\documentclass[aps,pre,amsmath,amssymb,twocolumn,nofootinbib]{revtex4}
\usepackage{pslatex}
\usepackage{graphicx}
\usepackage{dcolumn}
\usepackage{bm}
\usepackage{natbib}

\begin{document}
\title{The Cole-Cole Law for Critical Dynamics in Glass-Forming Liquids}

\author{M. Sperl}
\affiliation{Duke University, Department of Physics, Box 90305,
Durham, NC 27708, USA}

\date{\today}
\begin{abstract}

Within the mode-coupling theory (MCT) for glassy dynamics, the asymptotic 
low-frequency expansions for the dynamical susceptibilities at critical 
points are compared to the expansions for the dynamic moduli; this shows 
that the convergence properties of the two expansions can be quite 
different. In some parameter regions, the leading-order expansion formula 
for the modulus describes the solutions of the MCT equations of motion 
outside the transient regime successfully; at the same time, the leading- 
and next-to-leading order expansion formulas for the susceptibility fail. 
In these cases, one can derive a Cole-Cole law for the susceptibilities; 
and this law accounts for the dynamics for frequencies below the band of 
microscopic excitations and above the high-frequency part of the 
$\alpha$-peak. It is shown that this scenario explains the 
optical-Kerr-effect data measured for salol and benzophenone (BZP). For 
BZP it is inferred that the depolarized light-scattering spectra exhibit a 
wing for the $\alpha$-peak within the Gigahertz band. This wing results 
from the crossover of the von~Schweidler-law part of the $\alpha$-peak to 
the high-frequency part of the Cole-Cole peak; and this crossover can be 
described quantitatively by the leading-order formulas of MCT for the 
modulus.

\end{abstract}

\pacs{61.20.Lc, 82.70.Dd, 64.70.Pf}

\maketitle

\section{Introduction\label{sec:intro}}

During the past 15~years, several new spectrometers have been introduced 
for the study of the glassy dynamics of liquids. The evolution of this 
complex slow dynamics upon decreasing the temperature $T$ or increasing 
the density $\rho$ has been documented for many systems for times $t$, 
which exceed the natural time scale $t_\text{mic}$ for condensed-matter 
motions by three or more orders of magnitude. Similar progress has been 
made for molecular-dynamics simulations of liquid models. In parallel to 
these experimental activities, a theory for the evolution of glassy 
dynamics has been developed which is referred to as mode-coupling theory 
(MCT). This theory is based on regular equations of motion for a set of 
auto-correlation functions. The MCT equations lead to fold bifurcations 
for the correlators' long-time limits if some control parameter like $T$ 
reaches a critical value $T_c$; this bifurcation describes a transition 
from a liquid to an amorphous solid. The distance of the control parameter 
from the critical value, say $\varepsilon = (T_c-T)/T_c$, can be used for 
the discussion of the bifurcation dynamics as a small parameter. For 
$\varepsilon$ tending to zero and times increasing to infinity, it is 
possible to calculate asymptotic solutions of the MCT equations. The 
leading-order results provide a set of general formulas, which explain the 
qualitative features of the bifurcation scenario. Many fits of data with 
these general formulas have been studied in order to test the relevance of 
the MCT for the explanation of the experimental facts \cite{Goetze1999}.

A set of general MCT results, which is of main interest in this paper, 
concerns the critical dynamics. This dynamics is described by 
auto-correlation functions $\phi(t)$ for control parameters at the 
bifurcation point, say $T=T_c$. Equivalently, one can consider the 
corresponding loss spectra $\chi''(\omega)$. These are the products of the 
frequency $\omega$ and the Fourier-cosine transform of $\phi(t)$. The 
central asymptotic formula is specified by (1) a positive number $f^c$, 
which is called the plateau, (2) a positive amplitude, say $A$, and (3) 
the critical exponent $a$, obeying $0<a<0.396$, $\lim_{t\rightarrow\infty} 
t^a[\phi(t)-f^c] = A$. This formula is equivalent to 
$\lim_{\omega\rightarrow 0} \chi''(\omega)/\omega^a = \sin(\pi 
a/2)\Gamma(1-a)A$, with $\Gamma$ denoting the gamma function. For a given 
transition point, all correlators are specified by the same exponent $a$, 
but different critical points can differ in their value for $a$. The 
leading-order long-time result for the correlator describes a power-law 
decay: $\phi(t)-f^c \propto 1/t^a$. Equivalently, the leading-order result 
for the low-frequency critical loss spectrum is given by a power law 
variation $\chi''(\omega) \propto \omega^a$. For states near the 
transition point, $\phi(t)$ and $\chi''(\omega)$ can be replaced by their 
respective critical functions for short times, $t \ll t_\sigma$, or large 
frequencies, $\omega t_\sigma \gg 1$. The time scale $t_\sigma$ is the 
same for all correlators and diverges for states approaching the 
transition point. For $t \geqslant t_\sigma$ and $\omega t_\sigma 
\leqslant 1$, the correlators and spectra depend sensitively on 
$\varepsilon$; for shorter times, $t \ll t_\sigma$ and $\omega t_\sigma 
\gg 1$, the correlators and spectra depend on $\varepsilon$ smoothly. An 
$\omega^a$ spectrum was identified first for the glass-forming molten salt 
$0.4 Ca(NO_3)_2 0.6 K(NO_3)$ (CKN) in data obtained by neutron-scattering 
spectroscopy \cite{Knaak1988}. This system was also used to document for 
the first time the evolution of glassy dynamics within the full Giga-Hertz 
band \cite{Li1992}: Using depolarized light-scattering spectroscopy, a 
spectrum compatible with the $\omega^a$ law was found extending from 1GHz 
to 400GHz. The $t^{-a}$ decay in the time domain was measured first for 
density correlators by photon-correlation spectroscopy for a colloidal 
suspension of hard spheres \cite{Megen1993b} with the density as a control 
parameter.

Other glass-forming systems that can be studied experimentally are found 
in many van-der~Waals liquids. The natural time scale for inter-molecular 
vibrations is a picosecond. For normal liquid behavior, one expects 
correlations to decay to zero for times around some picoseconds. The 
normal-liquid excitation spectra extend from, say, 0.5THz to, say, 5THz. 
Glassy-dynamics spectra for several systems have been measured by 
depolarized light-scattering spectroscopy, and the data were shown to be 
consistent with the MCT bifurcation scenario. For example, spectra for 
toluene have been fitted successfully with leading-order asymptotic 
results for frequencies between 0.5GHz and 1THz and for temperatures 
decreasing from $T=T_c+140$K to $T_c+10$K; $T_c \approx 150$K 
\cite{Wiedersich2000}. For all these systems, the weight of the 
glassy-dynamics part of the spectra is large compared to the part of 
normal-liquid dynamics, i.e., the so-called $\alpha$-peak is large. In 
agreement with the MCT prediction for such situations, the amplitude $A$ 
for the $\omega^a$ spectrum is small, and the low-frequency contributions 
of the normal-liquid dynamics affect the region of the expected $\omega^a$ 
behavior of the spectra. As a result, no frequency interval has ever been 
identified for a van-der-Waals liquid, where an $\omega^a$ spectrum can be 
identified explicitly.

Torre \textit{et. al} \cite{Torre1998} have introduced optical-Kerr-effect 
(OKE) spectroscopy as a technique for the study of glassy dynamics. The 
measurement provides the response function $\chi(t) \propto 
-\partial_t\phi(t)$ for the same probing variable, which is studied in 
depolarized light-scattering experiments. The Fourier-sine transforms of 
$\chi(t)$ are proportional to the loss spectra $\chi''(\omega)$ mentioned 
in the preceding paragraph. The evolution of the glassy dynamics of 
m-toluidine was measured for temperatures decreasing from 295K to 250K. 
The response functions could be fitted well by the scaling-law results 
predicted by the leading-order asymptotic formulas for the MCT bifurcation 
\cite{Torre2000}. The analysis implies a critical temperature $T_c$ near 
220K and a critical exponent $a$ near 0.3. However, lowering the 
temperature to 225K, the critical power-law decay $\chi(t) \propto 
1/t^{1+a}$ was not observed \cite{Ricci2002}. The negative slope of the 
measured $\log\chi$-versus-$\log t$ curve is not $1+a$, rather it is a 
number smaller than unity, say $1-b'$. Decay laws $\chi(t) \propto 
1/t^{1-b'}$ with exponents $b'$ around 0.2 have been identified by Cang 
\textit{et. al} \cite{Cang2003c} for a number of other van-der-Waals 
liquids. The OKE response of salol was studied by Hinze \textit{et. al}, 
and the data were shown to be consistent with the known MCT scaling-law 
formulas for the temperature decreasing from 340K to 266K 
\cite{Hinze2000b}. The glassy response for $T = 257$K was measured with an 
impressive accuracy for times increasing up to 0.5~$\mu$s; and the 
dynamics for $t >$ 100ps displays the behavior expected from the 
leading-order asymptotic results of MCT. However, the glassy dynamics for 
2ps $<t<$ 20ps manifests itself by a $\chi(t) \propto 1/t$ decay. The 
corresponding correlator shows a logarithmic time dependence. All OKE 
response functions measured so far demonstrate a glassy dynamics that 
cannot be described by the general leading-order asymptotic formulas for 
the MCT bifurcation in the regime $t \lesssim 30$ps. In a recent 
alternative approach \cite{Berthier2005} is was possible to fit some of 
the OKE data for $t > 30$ps.

It was argued recently that the new facets of glassy dynamics discovered 
by OKE spectroscopy \cite{Ricci2002,Cang2003c,Hinze2000b} can be 
understood as generic implications of the Cole-Cole law for the critical 
dynamics \cite{Goetze2004}. In the following, this statement shall be 
explained in detail. Section~\ref{sec:general_MCT} summarizes the 
equations of motion and the known scaling-law results of MCT. It is 
explained in Sec.~\ref{sec:crit} that an asymptotic expansion of the 
modulus can have a much larger range of validity than the expansion for 
the susceptibility; the Cole-Cole law -- introduced in 1941 as empirical 
law \cite{Cole1941} -- is derived in full generality from the microscopic 
equations of motion. The relevance of these results is demonstrated for a 
schematic model in Sec.~\ref{sec:appl} using parameter values that 
describe the mentioned OKE data. The interplay between the Cole-Cole peak 
and the $\alpha$-peak is investigated in Sec.\ref{sec:alphabeta} where for 
specific parameter values the $\alpha$-peak displays a wing. 
Section~\ref{sec:sum} presents a conclusion.

\section{Essential MCT Formulas\label{sec:general_MCT}}
\subsection{Equations of Motion\label{subsec:basic_eqs}}

Within the basic version of MCT, the dynamics of the system is described 
by $M$ correlators $\phi_q(t)$, $q=1,\dots, M$. These are real and even 
functions of the time $t$, which obey the initial conditions $\phi_q(t=0) 
= 1$, $\partial_t\phi_q(t=0)=0$. The corresponding set of normalized 
response functions is given by

\begin{subequations}\label{eq:susc}
\begin{equation}\label{eq:susc:der}
\chi_q(t) = -\partial_t \phi_q(t)\,.
\end{equation}
Laplace transforms map functions from the time domain, say $F(t)$, in the 
frequency domain. They shall be used with the convention 
$\mathrm{LT}[F(t)](z) = i \int_0^\infty\,dt\,\exp[izt] F(t)$, $z = \omega 
+ i 0$. One gets $\mathrm{LT}[\chi_q(t)](z) = [1+\omega \phi_q(\omega)]$, 
where $\phi_q(\omega) = \mathrm{LT}[\phi_q(t)](z)$. The normalized 
dynamical susceptibilities are given by

\begin{equation}\label{eq:susc:FDT}
\chi_q(\omega) = 1+\omega\phi_q(\omega)\,.
\end{equation}
\end{subequations}
The loss spectra $\chi''_q(\omega) = \mathrm{Im}\, \chi_q(\omega)$ are 
related trivially to the fluctuation spectra $\phi_q''(\omega) = 
\mathrm{Im}\, \phi_q(\omega)$: $\chi_q''(\omega) = \omega 
\phi_q''(\omega)$.

The Zwanzig-Mori formalism provides a fraction representation of 
$\phi_q(\omega)$ in terms of a fluctuating-force correlator $M_q(\omega)$: 
$\phi_q(\omega) = -1/\{ \omega - \Omega_q^2 / [\omega + M_q(\omega)] \}$. 
The positive frequency $\Omega_q$ quantifies the initial decay of the 
correlator $\phi_q(t) = 1 - [\Omega_q t]^2/2 + {\cal O}(t^3)$. Within MCT, 
a white-noise term $\nu_q$ is split off from the kernel $M_q(\omega)$; the 
remainder is represented in terms of a dimensionless function $m_q(t)$: 
$M_q(\omega) = i\nu_q + \Omega_q^2 m_q(\omega)$; $\nu_q\geqslant 0$. Here, 
$m_q(t)$ and $m_q(\omega)$ are related by Laplace-transformation. The 
fraction representation is equivalent to the equations of motion

\begin{subequations}\label{eq:EOM}
\begin{equation}\label{eq:EOM:int}
\begin{split}
\partial_t^2 \phi_q(t) +& \nu_q\partial_t\phi_q(t)\hfill \\
+& \Omega_q^2 \left[
\phi_q(t)+\int_0^t\,dt'\,m_q(t-t')\partial_{t'}\phi_q(t')
\right] = 0\,.
\end{split}
\end{equation}
The essential approximation in MCT is the expression of $m_q(t)$ as a 
polynomial 
${\cal F}_q$ of the correlators: 
\begin{equation}\label{eq:EOM:kernel}
m_q(t) = {\cal F}_q[\phi_1(t), \dots, \phi_M(t)]\,.
\end{equation}
\end{subequations}
There is no monomial contribution of order zero. The coefficients of the 
polynomial are called mode-coupling coefficients. They are the coupling 
constants of the theory and must not be negative. Within the microscopic 
theory, the polynomials are of second order and the coefficients are given 
by the equilibrium structure functions, which in turn are smooth functions 
of control parameters like the temperature T for the states considered.

At the generic transition mentioned in the preceding section, the 
correlator's long-time limits depend singularly on the distance parameter 
$\varepsilon$. For $\varepsilon < 0$, fluctuations disappear for long 
times, $\phi_q(t\rightarrow\infty) = 0$. For $\varepsilon \geqslant 0$, 
$\phi_q(t \rightarrow \infty) = f_q$, $0<f_q<1$. $q = 1,\dots, M$; the 
fluctuations arrest. Within the microscopic version of MCT, the arrested 
part $f_q$ has the meaning of the Debye-Waller factor of the solid 
amorphous state. The leading-order variation with changes of $\varepsilon$ 
for the arrested part is given by $f_q - f^c_q \propto 
\sqrt{\varepsilon}$, $\varepsilon \rightarrow 0^+$, $f_q^c > 0$, 
$q=1,\dots, M$. The regularity of the MCT equations implies the following. 
For $\varepsilon$ tending to arbitrarily small values, there appears an 
arbitrarily large time interval, where $\phi_q(t)$ is arbitrarily close to 
$f^c_q$. This critical arrested part $f^c_q$ has the meaning of a plateau 
for the $\phi_q(t)$-versus-$\log t$ curves for states near the transition. 
The corresponding plateau for the force correlators is $m_q(t \rightarrow 
\infty) = f^{m\,c}_q = {\cal F}^c_q[f^c_1, \dots, f^c_M]$. At the 
transition, $\phi_q(\omega)$ and $m_q(\omega)$ exhibit poles 
$-f^c_q/\omega$ and $-f^{m\,c}_q/\omega$, respectively. Hence, there is a 
region of small $|\varepsilon|$ and small $\omega$, where $\omega+i\nu_q$ 
can be neglected compared to $\Omega_q^2 m_q(\omega)$. This region is the 
one for the MCT glassy dynamics. The fraction representation of the 
correlators simplifies to $\phi_q(\omega) = -1/[\omega - 1/m_q(\omega)]$. 
This formula can also be noted as

\begin{subequations}\label{eq:MCTlong} 
\begin{equation}\label{eq:MCTlong:m} 
\omega m_q(\omega) = \omega \phi_q(\omega)/\left[ 1 + \omega\phi_q(\omega) 
\right]\,. 
\end{equation} 
Equation~(\ref{eq:susc:FDT}) yields the equivalent expression for the 
dynamical susceptibility:

\begin{equation}\label{eq:MCTlong:chi} 
\chi_q(\omega) = 1/\left[ 1-\omega m_q(\omega)  \right]\,. 
\end{equation} \end{subequations} 
Within the regime of glassy dynamics, $\left[1-\omega m_q(\omega)\right]$ 
has the meaning of a modulus for the response described by $\chi_q(t)$. 
The pair of Eqs.~(\ref{eq:EOM:kernel}, \ref{eq:MCTlong:m}) can fix the 
solution only up to some overall time scale $t_0$. The latter is 
determined by matching of the transient dynamics with the glassy dynamics. 
For further details and for a list of original papers, the reader can 
consult Ref.~\cite{Franosch1997}.

\subsection{Scaling Laws\label{subsec:scaling_laws}}

There is a straight-forward recipe to calculate from the coupling 
coefficients at the transition point and from the critical arrested parts 
$f^c_q$ a number $\lambda$, $1/2 \leqslant \lambda < 1$. It is called the 
exponent parameter for the chosen transition point of the model under 
discussion. It fixes the critical exponent $a$, mentioned in 
Sec.~\ref{sec:intro}. It fixes a further exponent $b$, $0<b\leqslant 1$, 
which is called the von~Schweidler exponent. The equation for the two 
exponents reads $\Gamma(1-a)^2/\Gamma(1-2a) = \lambda = 
\Gamma(1+b)^2/\Gamma(1+2b)$. Parameter $\lambda$ also specifies a pair of 
equations for a pair of functions $g_\pm(\hat{t})$, which are defined for 
$\hat{t}>0$. The equations read $\pm 1 + \lambda g_\pm(\hat{t})^2 = 
(d/d\hat{t}) \int_0^{\hat{t}}\,d\hat{t}\, g_\pm(\hat{t}-\hat{t}') 
g_\pm(\hat{t}')$, and they have to be solved with the initial condition 
$\lim_{\hat{t}\rightarrow 0} \hat{t}^a g_\pm(\hat{t}) = 1$. Up to 
corrections of order $\hat{t}^a$, one gets for small $\hat{t}$:

\begin{subequations}\label{eq:gpm}
\begin{equation}\label{eq:gpm:ta}
g_\pm(\hat{t}\ll 1) = 1/\hat{t}^a\,.
\end{equation}
The function $g_+(\hat{t})$ approaches its long-time limit exponentially, 
$g_+(\hat{t}\gg 1) = 1/\sqrt{1-\lambda}$.
The function $g_-(\hat{t})$ exhibits a power-law divergence for large 
$\hat{t}$. Up to corrections of order $1/\hat{t}^b$, one gets

\begin{equation}\label{eq:gpm:B}
g_-(\hat{t}\gg 1) = -B\hat{t}^b\,.
\end{equation}
\end{subequations}
There are tables allowing the determination of $a$, $b$ and $B$ from a 
given $\lambda$. There are also tables to determine $g_\pm(\hat{t})$ with 
an accuracy sufficient for all practical purposes \cite{Goetze1990}.

The functions $g_\pm(\hat{t})$ are the shape functions for the first 
scaling law of MCT. For $\varepsilon$ tending to zero, there appears a 
time interval of diverging length, within which $\Delta_q(t) = |\phi_q(t) 
- f^c_q|$ is arbitrary small. The leading-order solution for the small 
parameter $\Delta_q(t)$ yields the first scaling-law. The solution 
assumes the form

\begin{subequations}\label{eq:phi_}
\begin{equation}\label{eq:phi_gpm}
\phi_q(t) = f^c_q + h_q \sqrt{|\sigma|} g_\pm(t/t_\sigma)\,,\quad 
\varepsilon \gtrless 0\,.
\end{equation}
The amplitudes $h_q > 0$, $q = 1, \dots, M$ are calculated from the 
mode-coupling coefficients at the critical point. The separation parameter 
$\sigma$ is defined similarly: it is a smooth function of the control 
parameters and can be linearized close to the transition point, $\sigma = 
C\varepsilon + {\cal O}(\varepsilon^2)$, with a positive coefficient $C$ 
that depends on the chosen control parameter. The first critical time 
scale $t_\sigma$ is given by the critical exponent $a$, the separation 
parameter $\sigma$, and the time scale $t_0$, which is defined by the 
short-time dynamics:

\begin{equation}\label{eq:sigma:tsigma}
t_\sigma = t_0/|\sigma|^\delta\,,\quad \delta = 1/(2a)\,.
\end{equation} \end{subequations}
The strong control-parameter dependence of the correlators near the 
plateau is described solely by that of the correlation scale 
$\sqrt{|\sigma|}$ and of the time scale $t_\sigma$.

From Eqs.~(\ref{eq:gpm:ta},\ref{eq:phi_gpm},\ref{eq:sigma:tsigma}), one 
gets the leading-order asymptotic law for the decay of the critical 
correlator discussed in Sec.~\ref{sec:intro}: $\phi_q(t) - f^c_q = h_q 
(t_0/t)^a$. From Eqs.~(\ref{eq:gpm:B}, \ref{eq:phi_gpm}, 
\ref{eq:sigma:tsigma}) one gets the von~Schweidler law for the decay of 
the liquid correlators below the plateau:

\begin{equation}\label{eq:vS:phi}
\phi_q(t) = f^c_q + h_q (t/t_{\sigma'})^b\,,\quad t_\sigma \ll t \ll 
t_{\sigma'}\,,\quad \varepsilon < 0 \,.
\end{equation}
Here, the second critical time scale of MCT reads $t_{\sigma'} = t_0/[ 
B^{1/b} |\sigma|^\gamma]$, with $\gamma = 1/(2a) + 1/(2b) \,. $ Up to 
errors of order $|\varepsilon|$, the decay of the liquid correlators below 
the plateau is described by the second scaling law of MCT

\begin{equation}\label{eq:alpha}
\phi_q(t) = \tilde{\phi}_q(t/t'_\sigma) \,,\quad
t'_\sigma \ll t\,,\quad \varepsilon <0\,.
\end{equation}
Here, $\tilde{\phi}_q(\tilde{t})$ is an $\varepsilon$-independent shape 
function. Its initial part is given by von~Schweidler's law: 
$\tilde{\phi}_q(\tilde{t}\ll 1) = f^c_q - h_q \tilde{t}^b$. Corrections of 
order $\sqrt{|\varepsilon|}$ modify the formula (\ref{eq:alpha}) for the 
below-plateau decay in a regime where $\phi_q(t)$ is close to the plateau. 
These corrections are described by Eq.~(\ref{eq:phi_gpm}) for $t \geqslant 
t_\sigma$. The below-plateau decay is referred to as $\alpha$-process, and 
Eq.~(\ref{eq:alpha}) formulates the superposition principle. Details of 
the derivation of the cited results, references to the original work, and 
a comprehensive demonstration for the hard-sphere system can be found in 
Refs.~\cite{Franosch1997,Fuchs1998}.

\section{Critical Dynamics and Cole-Cole law\label{sec:crit}}

In this section, correlators and susceptibilities shall be discussed for 
states at the transition point, say $T=T_c$. Focusing on the range of 
validity, the essential formulas are analyzed for the correlators in 
\ref{sec:crit:corr}, and for the susceptibilities in \ref{sec:crit:susc}. 
The Cole-Cole law is derived in \ref{sec:crit:cole}.

\subsection{Power-Law Solution for Correlation 
Functions\label{sec:crit:corr}}

The leading-order scaling-law formulas~(\ref{eq:gpm:ta},\ref{eq:phi_gpm}) 
yield for the critical correlators

\begin{equation}\label{eq:crit:corr}
\phi_q(t) = f^c_q + h_q (t_0/t)^a\,.
\end{equation}
A first question, whose answer is not implied by the results cited in 
Sec.~\ref{subsec:scaling_laws}, concerns the range of validity of 
Eq.~(\ref{eq:crit:corr}) for short times. Let us define an onset time 
$t^*_q$ for the power law by the request, that Eq.~(\ref{eq:crit:corr}) 
describes $\phi_q(t)- f^c_q$ within a relative error of, say, 10\%, for $t 
\geqslant t^*_q$. For normal-liquid dynamics, one would expect the 
correlators to decay to zero for times around $t_\text{mic}$. A second 
question to be discussed is: How can one describe the critical correlators 
in cases where $t^*_q$ exceeds $t_\text{mic}$, i.e., when there is a gap 
between the end of the transient regime and the onset of the critical 
power law?

The asymptotic solution~(\ref{eq:crit:corr}) can be extended to an 
asymptotic series expansion in powers of $t^{-a}$. The result up to the 
next-to-leading term shall be noted as

\begin{subequations}\label{eq:critcorr}
\begin{equation}\label{eq:critcorr:phi}
\phi_q(t) = f^c_q + h_q (t_0/t)^a \left[ 1+ \hat{K}_q (t_0/t)^a \right]\,.
\end{equation}
Here, remainders which are given by $(t_0/t)^{3a}$ times some power of 
$\ln(t/t_0)$ are dropped. The Tauberian theorem yields an equivalent 
formula in the frequency domain: 

\begin{equation}\label{eq:critcorr:chi}
\begin{split}
\omega\phi_q(\omega) = -f^c_q - &\Gamma(1-a) h_q (-i\omega t_0)^a\\
 &- \Gamma(1-2 a) h_q \hat{K}_q (-i\omega t_0)^{2a}\,. 
\end{split}\end{equation} 
\end{subequations} 
In this formula, terms are dropped, which are proportional to 
$\omega^{3a}$ times some power of $\ln\omega$. The preceding two formulas 
are the starting point for the following derivations. It is a 
straight-forward procedure to calculate the correction amplitude 
$\hat{K}_q$ from the mode-coupling coefficients at the critical point 
$T=T_c$ \cite{Franosch1997}. From Eq.~(\ref{eq:critcorr:phi}), one can 
estimate an onset time of $t^*_q/t_0 = [10|\hat{K}_q|]^\frac{1}{a}$. This 
is an estimate based on the assumption that higher-order expansion terms 
in Eq.~(\ref{eq:critcorr:phi}) do not influence the results seriously for 
$t \geqslant t^*_q$. Typically for many systems, the critical exponent $a$ 
is around 0.3, and $1/a \gtrsim 3$. Hence, $t^*_q$ depends sensitively on 
the correction amplitude $\hat{K}_q$. As a result, $t^*_q$ can vary 
considerably for different $q$. As a relevant example, let us cite the 
results for the density-fluctuation correlators of a system of hard 
spheres of diameter $d$ \cite{Franosch1997}: The time $t_1^*$ referring to 
a wave-number $q_1d = 7.0$ exceeds the time $t_2^*$ for $q_2d = 10.6$ by a 
factor of around 100.

\subsection{Power-Law Solution for the 
Susceptibilities\label{sec:crit:susc}}

From Eqs.~(\ref{eq:susc:der}, \ref{eq:crit:corr}), one obtains the 
long-time result for the response functions up to leading-order 
corrections,
\begin{equation}\label{eq:chit0}
\chi(t) t_0 = a h_q (t_0/t)^{1+a}\,.
\end{equation}
According to Eq.~(\ref{eq:critcorr:phi}), the leading order result in 
Eq.~(\ref{eq:crit:corr}) is valid for $t \geqslant t_q^*$, but the onset 
time for Eq.~(\ref{eq:chit0}) is later, $t \geqslant t^*_q 2^{1/a}$, where 
typically $2^{1/a} \approx 10$. Hence, the detection of the critical 
dynamics in its leading asymptotic form is more difficult for the response 
functions than for the correlators.

From Eq.~(\ref{eq:critcorr:chi}), one arrives at the expansion formula for 
the absorptive part of the dynamical susceptibility 

\begin{subequations}\label{eq:chi_power}
\begin{equation}\label{eq:chi_power:chi}\begin{split}
\chi_q''(\omega) = \left[ \Gamma(1-a) \sin\left( \pi a/2 \right) \right] 
&h_q (\omega t_0)^a \\&\times
\left[ 1+ k_a \hat{K}_q (\omega t_0)^a \right]\,,
\end{split}\end{equation}

\begin{equation}\label{eq:chi_power:ka}
k_a =  2 \Gamma(1-a) \cos\left( \pi a/2 \right)/\lambda\,.
\end{equation}
The leading-order power-law result reads

\begin{equation}\label{eq:chi_power:leading}
\chi_q''(\omega) = \left[ \Gamma(1-a) \sin\left( \pi a/2 \right) \right] 
h_q (\omega t_0)^a\,,
\end{equation}
\end{subequations}
which has an onset frequency of $\omega_q^* = 1/(t^*_q k_a^{1/a})$; for 
$\omega \leqslant \omega_q^*$, the leading-order result in 
Eq.~(\ref{eq:chi_power:leading}) describes the critical spectrum with an 
error smaller than 10\%. If $\lambda$ decreases from 1 to $1/2$, $k_a$ 
increases from 2 to near 5. For $\lambda$ near a typical value of 0.7, 
$k_a$ is above 3, and the onset frequency for the $\omega^a$-law is about 
30 times smaller than $1/t^*_q$. Therefore, the detection of the critical 
power-law in the loss spectra is even more difficult then in the response 
function.

Including the leading-correction terms in the asymptotic formulas, as 
noted in Eqs.~(\ref{eq:critcorr:phi}, \ref{eq:chi_power:chi}), is an 
obvious manner to extend the range of applicability of the analytic 
description of the dynamics. This is demonstrated comprehensively for the 
MCT for the hard-sphere system in Refs.~\cite{Franosch1997, Fuchs1998}. 
But there are cases, where this procedure does not lead to satisfactory 
results. The mean-squared displacement is an example, where the 
description of the increase towards the plateau by the analog of 
Eq.~(\ref{eq:critcorr:phi}) cannot account for the glassy dynamics 
\cite{Fuchs1998}.

\subsection{The Cole-Cole Law\label{sec:crit:cole}}

Equation~(\ref{eq:critcorr:chi}) is an asymptotic expansion in terms of 
powers of the small quantity $\xi = (-i\omega t_0)^a$ that holds up to 
errors of $\xi^3$. Substitution of this expansion into 
Eq.~(\ref{eq:MCTlong:m}) provides an analogous expansion for $\omega 
m_q(\omega)$. Let us indicate the non-trivial parts of the coefficients by 
a superscript $m$:

\begin{equation}\label{eq:mqw}\begin{split}
-\omega m_q(\omega) = f_q^{m\,c} +& \Gamma(1-a) h^m_q (-i \omega t_0)^a \\
&+ \Gamma(1-2a) h^m_q \hat{K}^m_q (-i \omega t_0)^{2a}\,.
\end{split}\end{equation}
Comparing coefficients of equal powers of $\xi$, one finds:

\begin{subequations}\label{eq:cmq}
\begin{equation}\label{eq:fqmc}
f_q^{m\,c} = f_q^c/(1-f_q^c)\,,\quad f_q^c =  f_q^{m\,c} /(1 +  
f_q^{m\,c} ) \,.
\end{equation}
\begin{equation}\label{eq:hmq}
h_q^m = h_q /(1-f_q^c)^2\,,\quad h_q = h_q^m/(1+ f_q^{m\,c} )^2\,.
\end{equation}
\begin{equation}\label{eq:cmq:cmq}\begin{split}
\hat{K}^m_q = \hat{K}_q + \lambda [h_q/(1-f_q^c)]\,,\quad\\
\hat{K}_q = \hat{K}^m_q -\lambda [h^m_q / (1 +  f_q^{m\,c} )]\,.
\end{split}
\end{equation}
\end{subequations}

Since the auto-correlation functions are normalized, $\phi_q(t=0) = 1$, 
one gets $f_q^c < 1$. The kernel $m_q(t=0)$ can have any positive value. 
Therefore, in principle, $f_q^{m\,c}$ can be any positive number. The 
different normalizations can be eliminated in the critical amplitudes by 
comparing the ratios $h_q/f_q^c$ and $h^m_q/f_q^{m\,c}$:

\begin{equation}\label{eq:hqfq}
[h^m_q/f^{m\,c}_q] = [h_q/f_q^c]/(1-f^c_q)\,.
\end{equation}
These ratios determine the relative amplitude of the dynamics around the 
plateau within the first scaling-law regime. Equation (\ref{eq:phi_gpm}) 
yields $(\phi_q(t)-f^c_q)/f^c_q = [h_q/f_q^c] \sqrt{|\sigma|} 
g_\pm(t/t_\sigma)$, and a corresponding identity holds for $m_q(t)$. From 
Eq.~(\ref{eq:hqfq}) one concludes that the relative amplitude of the 
first-scaling-law contribution is larger by $1/(1-f_q^c) > 1$ for the 
modulus than for the susceptibility; and $1/(1-f_q^c)$ increases with 
$f^c_q$, the $\alpha$-peak contribution of the loss spectrum. 
Consequently, within the range of validity of the first scaling law, it is 
easier to measure the first-scaling-law contribution for the modulus than 
that for the susceptibility.

The expansion~(\ref{eq:mqw}) can be substituted into 
Eq.~(\ref{eq:MCTlong:chi}) in order to obtain an asymptotic expansion for 
the inverse of the critical susceptibility, i.e., for the modulus. The 
result shall be noted in the form

\begin{equation}\label{eq:ccc}
\chi_q(\omega) = \chi_{0\,q}^{cc} /\left[ 1 + \left( 
-i\omega/\omega_q^{c}  \right)^a + \hat{K}^{cc}_q  \left( 
-i\omega/\omega_q^{c}  \right)^{2a}\right] \,.
\end{equation}
The expression $\chi_q(\omega)^{-1}$ is correct up to errors of the order 
$\omega^{3a}$. The three parameters specifying this formula can be 
expressed in terms of the coefficients in Eq.~(\ref{eq:mqw}) and, via 
Eqs.~(\ref{eq:cmq}), in terms of the coefficients 
specifying the correlators in Eq.~(\ref{eq:critcorr:phi}). One gets 
$\chi_{0\,q}^{cc} = 1/(1+f_q^{m\,c})$. This amplitude is the complement of 
the $\alpha$-peak strength $f^c_q$ of the normalized loss spectrum,

\begin{subequations}\label{eq:cccpar}
\begin{equation}\label{eq:cccpar:chi0}
\chi_{0\,q}^{cc} = 1- f^c_q\,.
\end{equation}
The characteristic frequency entering the new formula for the 
susceptibility, is given by $(\omega_q^{c}t_0)^a = (1 + f_q^{m\,c}) / 
\left( h^m_q \Gamma(1-a) \right)$ or by

\begin{equation}\label{eq:cccpar:wcc}
\omega_q^{c}t_0 = \left[ (1-f^c_q) / \left(h_q \Gamma(1-a)\right)  
\right]^{1/a}
\,.
\end{equation}
The correction amplitude reads 
$\hat{K}_q^{cc} = \hat{K}^m_q (1+f^{m\,c}_q)/(\lambda h^m_q)$,
which is equivalent to 
\begin{equation}\label{eq:cccpar:cqcc}
\hat{K}_q^{cc} = 1 + \left[(1-f^c_q)/h_q\right] \hat{K}_q / \lambda\,. 
\end{equation}
\end{subequations}

If the frequencies are so small that $|\hat{K}_q^{cc} 
(-i\omega/\omega_q^{c})| \ll 1$, Eq.~(\ref{eq:ccc}) simplifies to the 
transparent expression of the Cole-Cole law,

\begin{equation}\label{eq:ccleft:chi}
\chi_q(\omega) = \chi_{0\,q}^{cc}/\left[ 1 + (-i\omega/\omega_q^{c})^a 
\right]\,.
\end{equation}
The condition of validity for this formula means that the critical modulus 
can be described by the simple power law in the first line of 
Eq.~(\ref{eq:mqw}). The modulus spectrum obeys a formula analog to 
Eq.~(\ref{eq:chi_power:leading}): $\omega m_q''(\omega) = \Gamma(1-a) 
\sin\left(\pi a/2\right) h^m_q (\omega t_0)^a$. Using the 10\% criterion 
from above, the last term in Eq.~(\ref{eq:mqw}) can be dropped for 
frequencies $\omega t_0 \leqslant \left[ \lambda/\left( 10 \Gamma(1-a) 
|\hat{K}^m_q| \right) \right]^{1/a}$. The correlator corresponding to the 
loss spectrum in Eq.~(\ref{eq:ccleft:chi}) is given by the Mittag-Leffler 
function of index $a$: $M_a(x) = \sum_{n=0}^\infty x^n/\Gamma(1+n a)$,

\begin{equation}\label{eq:ML}
\phi_q(t) = f_q^c + \chi_{0\,q}^{cc} M_a\left[-(t\omega_q^{c})^a\right] 
\,.
\end{equation}
The Mittag-Leffler function can be calculated efficiently by Fourier-back 
transformation of $\chi''_q(\omega) / \omega$.

Equation~(\ref{eq:ccleft:chi}) describes a very broad peak for the loss 
spectrum $\chi_q''(\omega)$ with a maximum located at $\omega = 
\omega_q^{c}$. The spectrum is invariant under the interchange of 
$(\omega/\omega_q^{c})$ to $(\omega_q^{c}/\omega)$: the 
$\chi_q''(\omega)$-versus-$\log\omega$ curve is symmetric. The width of 
the peak increases strongly with decreasing exponent $a$. For $a \approx 
0.3$, $\omega$ has to increase by more than a factor of $10^4$ in order to 
scan the interval of frequency where $\chi_q''(\omega) \geqslant 
\chi_q''(\omega_q^{c}) / 2$. This susceptibility formula known as 
Cole-Cole law, Eq.~(\ref{eq:ccleft:chi}), was introduced as an empirical 
formula for dielectric loss spectra in glassy systems \cite{Cole1941}. In 
particular, broad peaks located above the $\alpha$ peaks -- often called 
$\beta$-peaks in this context -- have been fitted by it.

The Cole-Cole formula, Eq.~(\ref{eq:ccleft:chi}), exhibits simple limits 
for small and large frequencies. For $(\omega/\omega_q^{c})^a\ll 1$, the 
Cole-Cole susceptibility reproduces the general power-law for the loss 
spectrum, Eq.~(\ref{eq:chi_power:leading}), $\chi_q''(\omega) \propto 
\omega^a$. Similarly, the corresponding Mittag-Leffler correlator 
reproduces the general long-time asymptote for the response, 
Eq.~(\ref{eq:chit0}). For $(\omega/\omega_q^{c})^a \gg 1$, the Cole-Cole 
spectrum describes a critical spectrum which decreases with increasing 
frequency: $\chi_q''(\omega) \propto 1/\omega^a$. This is similar to a 
von~Schweidler-law spectrum. The corresponding Mittag-Leffler correlator 
decreases according to the law $\phi(t) -\text{const.} \propto
- t^a$. This yields a response function $\chi(t) \propto 1/t^x$ with $x = 
1-a <1$. Such behavior is consistent with the one detected by the OKE 
results for van-der Waals liquids \cite{Cang2003c}.

The crossover of the critical spectrum from the low-frequency wing to the 
high-frequency wing of the Cole-Cole loss peak has a counterpart for the 
Mittag-Leffler correlator that is most easily seen in a semilogarithmic 
plot. For $(t\omega_q^{c})^a \ll 1$, the $\left[ \phi(t) - \text{const.} 
\right]$-versus-$\log t$ curve is bent downward, as known for the 
von~Schweidler curves. For $(t\omega_q^{c})^a \gg 1$, there is the 
critical power-law decay which shows up as an upward-bent curve. Hence, 
the $\phi(t)$-versus-$\log t$ curve exhibits an inflection point. For an 
exponent $a$ around 0.3, the curve is nearly straight for an increase of 
$\log_{10} t$ by a factor of about 3. In this case, there is 
nearly-logarithmic decay of the critical correlator for a time variation 
over three orders of magnitude.

It depends on the size of $1/\omega_q^{c}$ relative to the microscopic 
time scale $t_\text{mic}$ which part of the Mittag-Leffler correlator 
dominates the critical dynamics in the time range of interest. The crucial 
question concerns the range of validity of the leading-order result for 
the susceptibility and for the modulus. These ranges are determined by the 
correction amplitudes. If $|\hat{K}_q|$ is large and $|\hat{K}^m_q|$ 
small, the Cole-Cole formula has a larger range of applicability than the 
general simple power-law formula~(\ref{eq:chi_power:leading}). But, if 
$|\hat{K}^m_q|$ is larger than $|\hat{K}_q|$, 
Eq.~(\ref{eq:chi_power:leading}) is a better approximation for the 
critical decay than the Cole-Cole spectrum. Since the brackets in 
Eqs.~(\ref{eq:cmq}) are positive, there are two obvious results. If 
$\hat{K}_q$ is positive, there holds $\hat{K}^m_q > \hat{K}_q$. In this 
case, the leading-order result for the susceptibility is a better 
description of the critical spectrum than the Cole-Cole spectrum. If 
$\hat{K}^m_q$ is negative, there holds $|\hat{K}^m_q| < |\hat{K}_q|$ and 
the application of the Cole-Cole law is superior to the one of the general 
leading-order result for the loss spectrum. For negative correction 
amplitudes $\hat{K}_q$, there is a trend for cancellation of the two 
contributions to $\hat{K}_q^m$ in Eq.~(\ref{eq:cmq:cmq}). It is a generic 
situation, that $|\hat{K}^m_q|$ is smaller than $|\hat{K}_q|$. Hence, for 
$\hat{K}_q < 0$, it is expected that the Cole-Cole law is a good 
description of the critical dynamics.

A detailed discussion of the MCT equations has shown that a large arrested 
part $f^c_q$ implies a large negative correction amplitude $\hat{K}_q$ for 
the critical decay \cite{Franosch1997}. In this case, the onset time 
$t^*_q$ may exceed the microscopic time scale $t_\text{mic}$ by several 
orders of magnitude. The time range for the applicability of the 
leading-order asymptotic description of the critical decay may be outside 
the accessible range of the existing spectrometers. But, this is the 
situation, where $|\hat{K}^m_q|$ may be so small that the Cole-Cole law or 
the Mittag-Leffler correlator can provide an adequate description of the 
critical dynamics.

\section{Application to Schematic Models\label{sec:appl}}

The preceding results shall be demonstrated in a simple schematic model 
for two different cases where the model parameters are adjusted to 
reproduce the data measured for benzophenone (BZP) and salol 
\cite{Hinze2000b,Cang2003c,Goetze2004}

The simplest MCT models deal with a single correlation function only. For 
these models, the formulas in the preceding sections simplify since the 
label $q$ can be dropped. The correlator, the critical arrested part, and 
the critical amplitude shall be denoted by $\phi(t)$, $f^c$, and $h$, 
respectively. The model is specified by the frequency $\Omega$ and the 
friction coefficient $\nu$ in the equation of motion~(\ref{eq:EOM:int}). 
Furthermore, there are the non-negative coefficients $v_l,\, l \geqslant 
1$, which specify the mode-coupling monomial of order $l$ for ${\cal F}$ 
in Eq.~(\ref{eq:EOM:kernel}). In applications for data descriptions, the 
specified numbers are considered as smooth functions of the physical 
control parameters.

The simplest model for the polynomial ${\cal F}$, which can reproduce all 
possible values for the exponent parameter $\lambda$, is given by the 
following formula for the fluctuating-force kernel

\begin{equation}\label{eq:mF12}
m(t) = v_1\phi(t) + v_2 \phi(t)^2\,.
\end{equation}
The two coupling constants $v_1$ and $v_2$ specify the state of the system 
by a point in the first quadrant of the $v_1$-$v_2$ plane, cf. lower inset 
in Figs.~\ref{fig:BZP} and \ref{fig:salol}. The points $(v_1^c, v_2^c)$ 
for generic fold bifurcations are located on a piece of a parabola. The 
position of the specific transition point can be characterized by the 
value for $\lambda$ \cite{Goetze1984}:

\begin{subequations}\label{eq:vc}
\begin{equation}\label{eq:vc:vc}
v_1^c = (2\lambda-1)/\lambda^2 \,,\; v_2^c = 1/\lambda^2 \,,\; f^c = 
1-\lambda \,,\; h=\lambda\,.
\end{equation}
The separation of some point $(v_1, v_2)$ from the transition point 
$(v_1^c, v_2^c)$, see upper insets Figs.~\ref{fig:BZP} and 
\ref{fig:salol}, is given by
\begin{equation}\label{eq:vc:sigma}
\sigma = \left[ \hat{v}_1 + \hat{v}_2 (1-\lambda) \right] \lambda 
(1-\lambda) \,,\quad \hat{v}_{1,\,2} = v_{1,\,2} - v^c_{1,\,2}
\,.
\end{equation}
\end{subequations}
The arrested part of the correlator for the glass state reads $\phi(t 
\rightarrow \infty) = h[\sigma/(1-\lambda)]^{1/2}+{\cal O}(\sigma^{3/2})$. 
The splitting of this value in an amplitude $h$ and a remainder is not 
unique. The cited value for the critical amplitude $h$ follows the 
conventions made in the preceding literature \cite{Franosch1997}. The 
general expression for the correction amplitude $\hat{K}$ for $M=1$ models 
\cite{Goetze1989c} is specialized easily to $\hat{K} = \kappa(x)$, where 
the functions $\kappa(x)$ is defined by

\begin{equation}\label{eq:kappa}
\kappa (x) = \frac{1}{2} \Gamma(1-x)^3/ \left[
\lambda \Gamma(1-3x) - \Gamma(1-x) \Gamma(1-2x)
\right] \,.
\end{equation}

Function $\kappa(x)$ increases monotonically with increasing $\lambda$: it 
decreases with increasing $x$. For $\lambda = 1/2$, i.e., for $a = 
0.395\dots$, one gets $\kappa(a) = -0.169\dots$ . For $\lambda$ 
approaching 1, $a$ tends to zero and $\kappa(a)$ diverges. For $a = 1/3$, 
$\lambda = 0.684\dots$, the correction amplitude vanishes. All 
higher-order corrections for the critical decay outside the transient 
regime vanish as well for this special value of $\lambda$ 
\cite{Fuchs1999b}. Therefore, for $\lambda$ near 0.7, the simple power-law 
formulas of the leading-order asymptotic-expansion theory like 
Eqs.~(\ref{eq:crit:corr}, \ref{eq:chit0}, \ref{eq:chi_power:leading}) 
describe the critical dynamics very well.

For the specified model, there holds $f^c\leqslant 1/2$. Hence, this model 
cannot be used to describe glassy-dynamics data for systems where the 
$\alpha$-peak loss spectra have a weight $f^c$ which exceeds 50\% of the 
total weight. Moreover, the relation between the exponent parameter 
$\lambda$ and the critical arrested part $f^c$, as formulated by 
Eq.~(\ref{eq:vc:vc}), is an artifact of the model. The minimum requirement 
for a schematic model for data analysis is the freedom to adjust $\lambda$ 
and $f^c$ independently. This goal can be achieved by introducing a second 
correlator. The first correlator $\phi_{q=1}(t) = \phi(t)$ is used as a 
caricature of the density-fluctuation dynamics. It provides the exponent 
parameter. The second one describes the dynamics of some probing variable, 
say, $A$. The correlator shall be denoted by $\phi_{q=2}(t) = \phi_A(t)$. 
All quantities referring to this correlator shall be indicated by an index 
$A$. The equation of motion~(\ref{eq:EOM:int}) is specified by $\Omega_A > 
0$, $\nu_A \geqslant 0$, and a kernel $m_A(t)$. The two frequencies 
quantify the transient dynamics. The kernel is a polynomial 
(\ref{eq:EOM:kernel}) of the two correlators involved. The simplest model 
describing the coupling of the probing variable to the density-fluctuation 
is given by \cite{Sjoegren1986}

\begin{equation}\label{eq:mA}
m_A(t) = v_A \phi(t) \phi_A(t)\,.
\end{equation}
The coupling of the probing variable to the density fluctuations is 
quantified by $v_A >0$. Since there is no influence of the dynamics of the 
probing variable on the density dynamics, the same scaling-law function 
$g_\pm(t/t_\sigma)$, Eq.~(\ref{eq:phi_gpm}), describes the leading-order 
near-plateau dynamics of $\phi_A(t)$ as it does for that of $\phi(t)$. 
Notice in particular that the parameters $\Omega_A$, $\nu_A$, and $v_A$ do 
not modify the time scale $t_0$. In applications for data descriptions, 
the model parameters $\Omega_A$, $\nu_A$, and $v_A$ are to be considered 
as smooth functions of the physical control parameters, cf. lower insets 
in Figs.~\ref{fig:BZP} and \ref{fig:salol}.

The loss spectrum $\chi_A''(\omega)$ develops a $\beta$-peak upon 
increasing the mode-coupling coefficient $v_A$ \cite{Buchalla1988}. For 
this special case and in the limit of infinite $v_A$, the Cole-Cole 
susceptibility and the Mittag-Leffler correlator have been derived for the 
critical dynamics of the probing variable in Ref. \cite{Goetze1989c}. The 
model specified by Eqs.~(\ref{eq:mF12}, \ref{eq:mA}) was used repeatedly 
for the description of experimental data 
\cite{Singh1998,Ruffle1999,Brodin2002,Wiebel2002,Goetze2004,Cang2005}. 
Large data sets for the evolution of the glassy dynamics of propylene 
carbonate have been analyzed in Ref.~\cite{Goetze2000b}; results measured 
for different probing variables $A$ obtained by neutron-scattering, 
depolarized-light-scattering, and dielectric-loss spectroscopy have been 
fitted by a common first correlator $\phi(t)$. The different probes have 
been characterized by adjusting the parameters for the second correlator 
$\phi_A(t)$ only.

Up to order $(t_0/t)^{2a}$, the kernel $m_A(t)$ can be calculated from 
Eq.~(\ref{eq:mA}) substituting the expression~(\ref{eq:critcorr:phi}) for 
$\phi(t)$ and the analog expression for the second correlator,

\begin{equation}\label{eq:phiA}
\phi_A(t) = f_A^c + h_a (t_0/t)^a \left[ 1 + \hat{K}_A (t_0/t)^a \right] 
\,.
\end{equation}
Laplace transform yields $\omega m_A(\omega)$ as an asymptotic power 
series in the small parameter $\xi = (-i\omega t_0)^a$. The coefficients 
are linear functions of $f^c_A$, $h_A$, and $\hat{K}_A$. This is 
substituted on the left-hand side of Eq.~(\ref{eq:MCTlong:m}) for $q=2$. 
Expanding the right-hand side up to orders $\xi^2$, one can compare the 
coefficients in order to arrive at:

\begin{equation}\label{eq:fAhA}
f_A^c = 1- \left[ 1/ (f^c v_A) \right] \,,\quad
h_A = \lambda / \left( f^{c\,2} v_A\right)\,.
\end{equation}

The correction amplitude shall be noted in the convention of the general 
theory \cite{Franosch1997}:

\begin{subequations}\label{eq:cAKA}
\begin{equation}\label{eq:cAKA:cA}
\hat{K}_A = \kappa(a) + K_A\,,
\end{equation}

\begin{equation}\label{eq:cAKA:KA}
K_A = \lambda v_a \left[ \frac{1}{v_A(1-\lambda)} -\lambda \right] / 
\left[ v_A(1-\lambda) - 1 \right]\,.
\end{equation}
\end{subequations}
This leads to the parameters for the Cole-Cole susceptibility:

\begin{equation}\label{eq:chi0ccA}
\chi^{cc}_{0\,A} = 1/\left(f^c v_A\right) \,,\quad 
\omega^{c}_A t_0 = \left[\left(1-\lambda\right)/\left(\lambda\Gamma(1-a) 
\right)\right]^{1/a} \,.
\end{equation}
Remarkably, the comparison of Eq.~(\ref{eq:chi0ccA}) with the general 
expression in Eq.~(\ref{eq:cccpar:wcc}) shows, that the Cole-Cole 
frequency $\omega_A^{c}$ does not depend on the coupling coefficient 
$v_A$. $\omega_A^{c}t_0$ decreases with increasing $\lambda$ from 
0.37\dots for $\lambda=1/2$ to about $10^{-4}$ for $\lambda$ near 0.88. 
For $\lambda$ around 0.7, $\omega_A^{c}t_0$ is about 0.02.

\subsection{Critical Relaxation for a Small Cole-Cole 
Frequency\label{subsec:BZP}}

\begin{figure}
\includegraphics[width=\columnwidth]{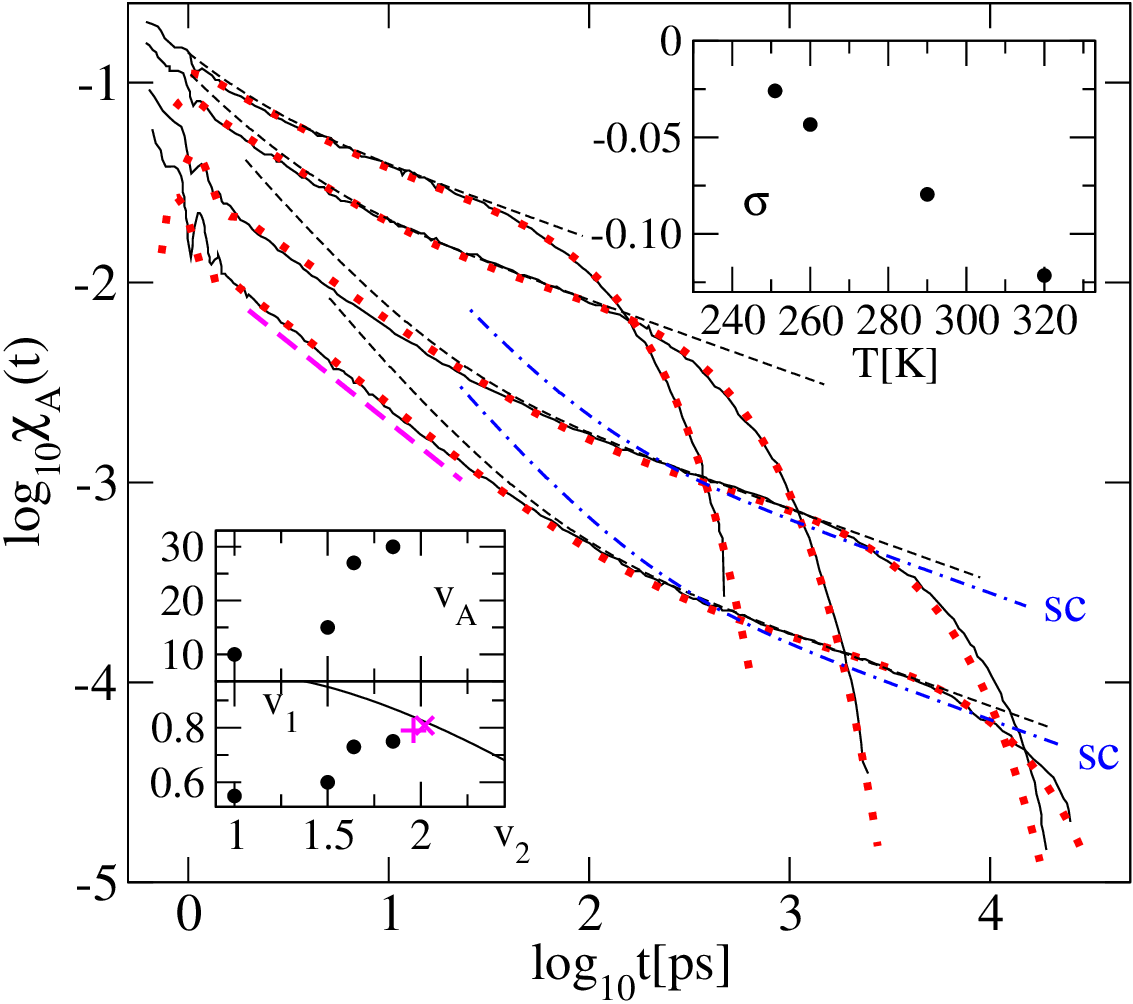}
\caption{\label{fig:BZP}OKE response functions measured for BZP for $T/K = 
251,\,260,\,290,\,320$ \cite{Cang2003c} (full lines from bottom to top)
and fits by the schematic-model functions $\chi_A(t)$ \cite{Goetze2004} 
(dotted lines). The straight dashed line has slope $-0.80$.
The dash-dotted lines labeled \textit{sc} are the 
approximations by the scaling functions Eq.~(\ref{eq:chiA_beta}), the thin 
dashed lines are fits with Eq.~(\ref{eq:chiA_beta}) for freely adjusted 
$s_\sigma$ and $t_\sigma$. The insets show the separation parameter 
$\sigma$ (right) and the fitting parameters (left); two additional state 
points ($+$ and $\times$) indicate extrapolations for which solutions are 
shown in Fig.~\ref{fig:BZPwing}.
}
\end{figure}

Figure~\ref{fig:BZP} reproduces OKE-response functions measured for 
benzophenone (BZP) \cite{Cang2003c} and fits to these data by the response 
$\chi_A(t)$ calculated for the model defined in the preceding section. The 
fit parameters for the mode-coupling coefficients $v_1$, $v_2$, and $v_A$ 
are specified in the lower inset. The analysis shall be done by 
anticipating a transition point for $\lambda=0.70$, which implies the 
exponents $a=0.33$ and $b=0.64$. The cross in the inset is close to this 
bifurcation point. The $\sigma$-versus-$T$ diagram is shown in the upper 
inset; and extrapolation to $\sigma=0$ suggests the critical temperature 
$T_c=235\text{K}$ with an estimated uncertainty of $\pm 5\text{K}$. The 
$251\text{K}$ result exhibits the expected von~Schweidler decay, 
$\log\chi(t) = \text{const.}-(1-b)\log t$ for times between about 0.3ns 
and about 6ns. Further details can be inferred from 
Ref.~\cite{Goetze2004}.

\begin{figure}
\includegraphics[width=\columnwidth]{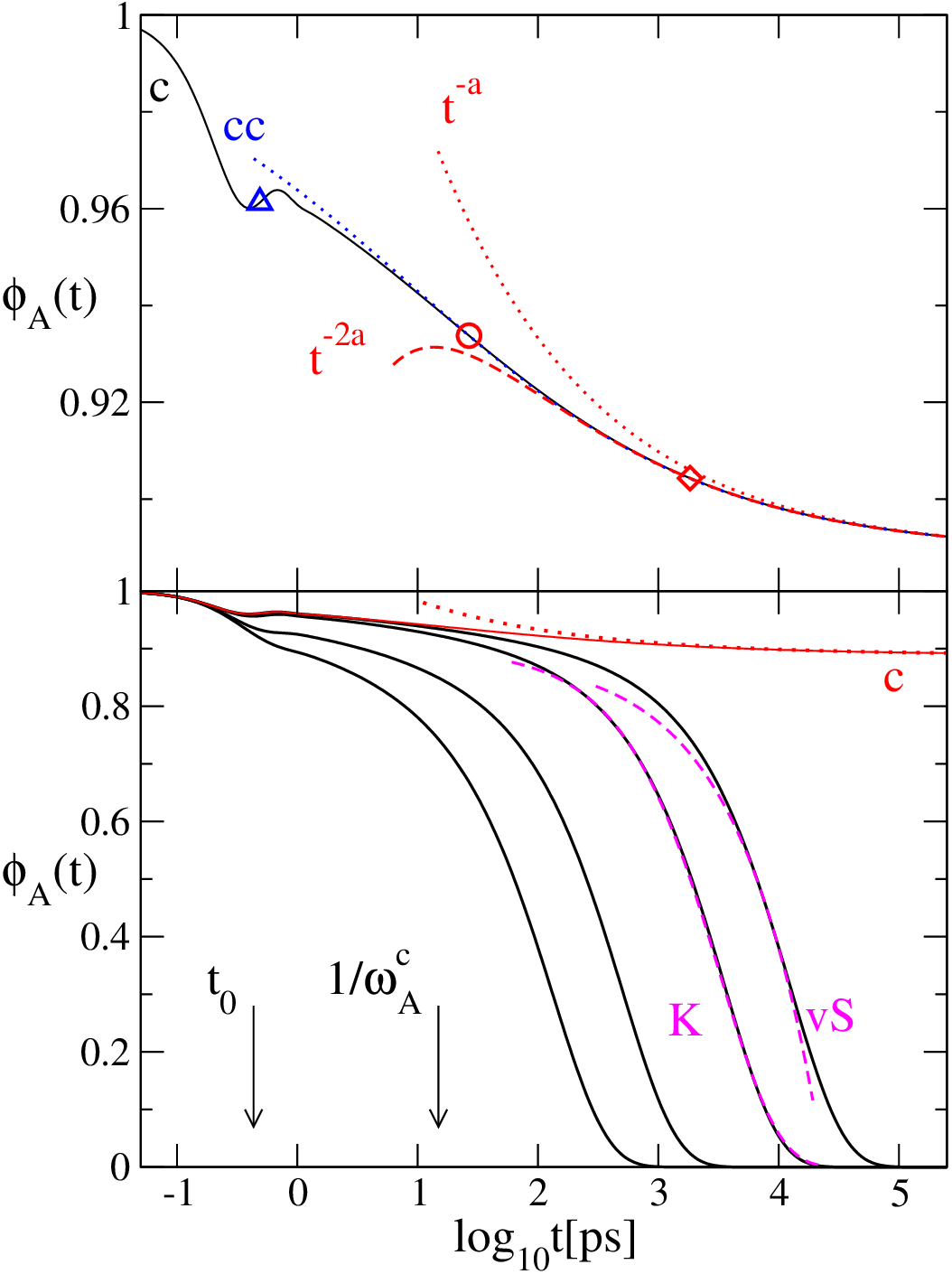}
\caption{\label{fig:BZPphiAt} Correlation functions for BZP.
The lower panel exhibits the correlators 
$\phi_A(t)$ underlying the fits in Fig.~\ref{fig:BZP} (heavy full lines 
from left to right for $T/K = 320,\,290,\,260,\,$ and $251$, 
respectively). The light full line with label \textit{c} shows the 
critical correlator calculated for $\lambda = 0.7$ and $v_A = 30$. The 
dotted line is the leading asymptotic law, Eq.~(\ref{eq:crit:corr}), with 
the time scale $t_0 = 0.3775$ps. The dashed line marked \textit{vS} shows 
a von~Schweidler law, Eq.~(\ref{eq:vS:phi}), with $b=0.64$ and a time 
scale adjusted to match the $T=251$K curve. The dashed line labeled 
\textit{K} is a fit by the Kohlrausch law, $\phi_A(t) = f_A^c 
\exp[-(t/\tau)^\beta],\,\beta=0.91$, with $\tau$ adjusted to match the 
$T=260$K curve. The times $t_0$ and $1/\omega^c_A = 15$ps are indicated by 
arrows. The upper panel shows the critical decay (\textit{c}) together 
with the leading-order ($t^{-a}$) and next-to-leading-order ($t^{-2a}$)
asymptotic power-law solution, Eq.~(\ref{eq:critcorr:phi}). The dotted 
curve labeled \textit{cc} displays the leading-order Cole-Cole solution, 
Eq.~(\ref{eq:ML}), with $\chi_0^{cc}=1/9$. The points of 10\% deviation of 
the approximations $t^{-a}$, $t^{-2a}$, and \textit{cc} from the critical 
decay are indicated by the diamond, circle, and triangle, respectively.
}
\end{figure}

\begin{figure}
\includegraphics[width=\columnwidth]{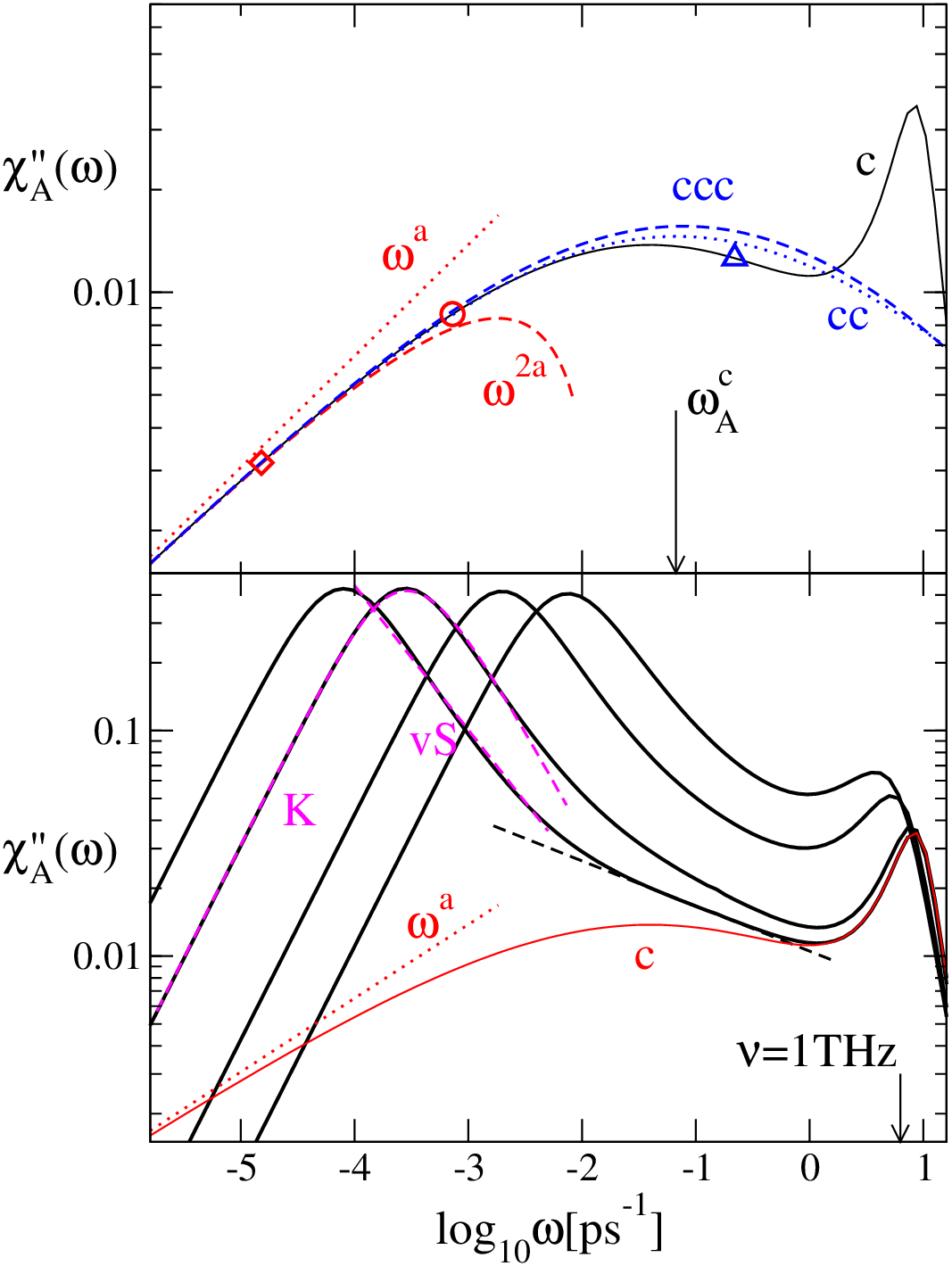}
\caption{\label{fig:BZPchi} Spectra for BZP.
The lower panel shows the susceptibility spectra for the correlators from 
the lower panel of Fig.~\ref{fig:BZPphiAt}. The leading-order critical 
spectrum, Eq.~(\ref{eq:chi_power:leading}), is the dotted straight
line marked by 
$\omega^a$. A dashed straight line with slope $-0.2$ is fitted to the 
$T=251$K spectrum to indicate an $\omega^{-b'}$ law for $b' = -0.2$. 
In the upper panel, the dashed curve marked $\omega^{2a}$ shows the  
approximation by Eq.~(\ref{eq:chi_power:chi}). The approximation by the 
Cole-Cole function (\textit{cc}), Eq.~(\ref{eq:ccleft:chi}), is shown 
dotted, the inclusion of the correction, Eq.~(\ref{eq:ccc}) yields the 
dashed curve labeled \textit{ccc} for $\omega_A^c = 67$ns$^{-1}$.
The points of 10\% deviation of the approximations $\omega^{-a}$, 
$\omega^{-2a}$, and \textit{cc} from the critical spectrum are indicated 
by the diamond, circle, and triangle, respectively. The frequencies 
$\omega_A^c$ and $\nu = \omega/(2\pi) = 1$THz are marked by arrows.
}
\end{figure}

The MCT results describe the measured evolution of the glassy dynamics 
adequately for times exceeding $t_\text{mic} = 1\text{ps}$ with two 
exceptions. The $T = 251\text{K}$ data exhibit a pronounced oscillation 
near $1.1\text{ps}$, while the calculated curve shows this oscillation 
near $0.9\text{ps}$. Furthermore, there are some deviations between the 
theoretical results and the data for the $T = 260\text{K}$ results for 
times around $10\text{ns}$. The four correlators $\phi_A(t)$ and the four 
loss spectra $\chi''_A(t)$, which are shown in the lower panels of 
Figs.~\ref{fig:BZPphiAt} and~\ref{fig:BZPchi}, respectively, are the MCT 
results corresponding to the MCT responses $\chi_A(t)$ shown in 
Fig.~\ref{fig:BZP}. The data fits have been calculated for fixed relative 
values of the four frequencies, which specify the transient dynamics: 
$\Omega_A = \Omega$, $\nu_A = \nu = 5\Omega$. The scale $\Omega$ for the 
fits is chosen different for different temperatures $T$ \cite{Goetze2004}. 
The curves in Fig.~\ref{fig:BZPphiAt} and~\ref{fig:BZPchi} are presented 
with a $T$-independent scale corresponding to that one used for the fit 
for $T = 251$K: $\Omega = 1.67$ps$^{-1}$. With the mentioned reservations, 
these curves can be considered as the measured quantities for the 
OKE-probing variable $A$. The full lines marked $c$ in 
Figs.~\ref{fig:BZPphiAt} and~\ref{fig:BZPchi} show the critical correlator 
$\phi_A^c(t)$ and the critical loss spectrum ${\chi''_A}^c(\omega)$, 
respectively, for the transition point specified by $\lambda = 0.70$ and 
$v_A = 30$. The correlator relaxes to the plateau $f_A^c = 8/9$ with a 
critical amplitude $h_A = 7/27$.

The second scaling law for the decay below the plateau of the correlators, 
Eq.~(\ref{eq:alpha}), implies a corresponding scaling law for the 
response:

\begin{equation}\label{eq:alpha_chi}
\chi_A(t) = \tilde{\chi}_A(t/t'_\sigma)/t'_\sigma\,.
\end{equation}
The control-parameter independent shape function reads 
$\tilde{\chi}_A(\tilde{t}) = -\partial\tilde{\phi}_A(\tilde{t})  / 
\partial\tilde{t}$. The reader can check this superposition principle for 
the results in Fig.~\ref{fig:BZP} as follows. The $\log 
\chi_A(t)$-versus-$\log t$ curve for $T = 320\text{K}$ and $t \geqslant 
2\text{ps}$ can be translated so that it collapses with the curves for 
$T/\text{K} = 290,\,260,\,$ and $251$ for times exceeding $4\text{ps}$, 
$90\text{ps}$, and $290\text{ps}$, respectively. Equivalently, the four 
loss spectra in Fig.~\ref{fig:BZPchi} for $\chi''_A(\omega) \geqslant 0.1$ 
are connected by the superposition principle: $\chi''_A(\omega) = 
\tilde{\chi}''_A(\omega t'_\sigma)$. The von~Schweidler law describes a 
part of the high-frequency wing of the loss peak as shown by the straight 
dashed line with label \textit{vS} for the $T = 251\text{K}$ curve. The 
Kohlrausch law for a stretching exponent $\beta = 0.91$ describes the 
upper part of the $\alpha$-peak as shown by the dashed line with label $K$ 
for the $T = 260\text{K}$ result.

There is an interval of times between the end of the transient dynamics 
and the start of the von~Schweidler decays that deals with the relaxation 
towards and through the plateau $f_A^c$. The dynamics causes the loss 
spectra for $\chi''_A(\omega) \leqslant 0.1$ and $\omega/(2\pi) < 0.1$THz 
shown in Fig.~\ref{fig:BZPchi}. The leading-order asymptotic solution of 
the MCT equations of motion deals with this part of the dynamics together 
with the von~Schweidler-law part by Eq.~(\ref{eq:phi_gpm}). This formula 
yields the first scaling law for the response

\begin{equation}\label{eq:chiA_beta}
\chi_A(t) = h_A s_\sigma \hat{\chi}(t/t_\sigma)\,;\quad
s_\sigma = \sqrt{|\sigma|}/t_\sigma\,.
\end{equation}

The shape function reads $\hat{\chi}(\hat{t}) = -\partial g_\pm(\hat{t}) / 
\partial\hat{t}$, $T \gtrless T_c$. For $T > T_c$, Eq.~(\ref{eq:gpm:B}) 
yields the von~Schweidler response for large rescaled times $\hat{t} = t / 
t_\sigma$: $\hat{\chi}(\hat{t} \gg 1) =B\cdot b / \hat{t}^x\,,\quad x = 1- 
b < 1$. For small rescaled times, Eq.~(\ref{eq:gpm:ta}) yields 
$\hat{\chi}(\hat{t} \ll 1) = a/ \hat{t}^x\,,\quad x = 1 + a > 1$. The 
crossover from one asymptote to the other occurs for times near 
$t_\sigma$. The light dashed lines in Fig.~\ref{fig:BZP} exhibit fits of 
the data by Eq.~(\ref{eq:chiA_beta}) for $\lambda = 0.70$. However, both 
scales $s_\sigma$ and $t_\sigma$ are adjusted with the aim to achieve a 
good match in the von~Schweidler-law regime. The $1/t^{1+a}$-law is not 
exhibited by the results in Fig.~\ref{fig:BZP}. For $T = 320\text{K}$, the 
crossover time $t_\sigma$ is located near $1\text{ps}$, i.e., it is within 
or close to the transient regime. For the other three temperatures and 
times within the interval $t_\text{mic} < t < t_\sigma$, the measured 
response and the calculated functions $\chi_A(t)$ are below the values of 
$h_A s_\sigma\hat{\chi}(t / t_\sigma)$.

The straight dashed line in Fig.~\ref{fig:BZP} has a slope of $-0.80$. It 
demonstrates a pseudo-von~Schweidler decay of the $T = 251\text{K}$ 
results for $2\text{ps} < t < 20\text{ps}$: $\phi_A(t) - \text{const.} 
\propto -t^{b'}\,,\quad b' = 0.20$. The $\phi_A(t)$-versus-$\log t$ curves 
in Fig.~\ref{fig:BZPphiAt} for $t > 2\text{ps}$ approach and cross the 
plateau as downward-bent curves before entering the von~Schweidler decay 
regime. There are no inflection points of the curves for some time near 
$t_\sigma$, as implied by Eq.~(\ref{eq:phi_gpm}).

The first scaling law describes a loss minimum of some value 
$\chi_\text{min}$ at some position $\omega_\text{min}$. The shape of the 
$\log \chi_A''(\omega)$-versus-$\log\omega$ curves is independent of the 
separation parameter $\sigma$ and the scales fix the position of the 
curve: $\chi_\text{min} \propto \sqrt{|\sigma|}$, $\omega_\text{min} 
\propto 1/t_\sigma$. The minimum is due to the crossover from the 
von~Schweidler wing of the $\alpha$-peak, $\chi''(\omega) \propto 1 / 
\omega^b$, to the power-law asymptote for the critical dynamics, 
$\chi''(\omega) \propto \omega^a$. For most practical purposes, the 
minimum can be approximated by the interpolation formula 
\cite{Sjoegren1990b}:

\begin{equation}\label{eq:interpol}
\chi''_A(\omega) / \chi_\text{min} = \left[
b \left(\omega/\omega_\text{min}\right)^a
+ a \left(\omega_\text{min}/\omega\right)^b
\right]/(a+b)\,.
\end{equation}
However, the loss spectra in the lower panel of Fig,~\ref{fig:BZPchi} do 
not exhibit loss minima, which can be described by the interpolation 
formula for the leading-order asymptotic result. The minima for 
frequencies $\omega$ near $1\text{ps}^{-1}$ do not depend sensitively on 
$\sigma$. With decreasing $|\sigma|$, the minimum position even shifts 
slightly upwards rather than downwards. With increasing frequency, the 
high-frequency part of the $\alpha$-peak exhibits a crossover from the 
von~Schweidler decay, $\chi''(\omega) \propto 1/\omega^b$, to a pseudo 
von~Schweidler decay, $\chi''(\omega) \propto 1/\omega^{b'}$. For $T = 
251$K, this high-frequency wing extends from $\omega \approx 
0.02\text{ps}^{-1}$ to about 0.4ps$^{-1}$ with $b' = 0.2$, as is 
demonstrated by the dashed straight line. The observed minimum is due to 
the crossover from the $1/\omega^{b'}$ wing to the spectral peak for the 
normal-liquid dynamics. The latter is located near $\nu= 1$THz.

The scaling-law fits, which are shown in Fig.~\ref{fig:BZP} by the light 
dashed lines, are misleading because the scales $s_\sigma$ and $t_\sigma$ 
have been adjusted freely. Calculating these scales from 
Eqs.~(\ref{eq:sigma:tsigma}, \ref{eq:vc:sigma}, \ref{eq:fAhA}), one 
obtains for $T = 251\text{K}$ and 260K the dash-dotted lines marked by 
$sc$. The two distance parameters $\varepsilon = (T_c -T) / T_c$ are -0.07 
and -0.11, respectively. These values are too large for a leading-order 
asymptotic formula to be applicable.

The correlator $\phi_A(t)$ for the lowest temperature under discussion is 
close to the critical one for $t$ up to about $30$ps, as is shown in the 
lower panel of Fig.~\ref{fig:BZPphiAt}. Hence, the $T = 251$K curves 
exhibit critical glassy dynamics for the times between 1ps and 30ps. From 
Eqs.~(\ref{eq:cAKA}) one obtains the correction amplitude $\hat{K}_A = 
-1.51$. This large value yields an onset time $t_A^*$ for the $t^{-a}$-law 
which is beyond $t = 10^3$ps as is marked by the diamond in the upper 
panel of the figure. Including the leading correction term yields the 
approximation by the dashed line denoted $t^{-2a}$. It improves the 
description of the critical decay so that it can be understood for times 
larger than about $t = 30$ps. But the expansion formula~(\ref{eq:phiA}) 
cannot be used to describe that part of the critical decay, which is 
measured for BZP and described by the solution of the two-component 
schematic MCT model.

Substituting the cited values for $\lambda$, $f_A^c$, and $h_A$ into 
Eqs.~(\ref{eq:fqmc}, \ref{eq:hmq}), one gets the plateau for the modulus,
$f_A^{m\,c} = 8$, and the critical amplitude $h_A^m = 21$. Notice that the 
relative strength of the first scaling law amplitude for the modulus is 
much larger than that for the correlator: $h_A^m/f_A^{m\,c} = 2.6$ versus 
$h_A/f_A^c = 0.29$. From Eq.~(\ref{eq:cmq:cmq}), one obtains the 
correction amplitude $\hat{K}_A^m = 0.124$. As expected from the 
discussions of 
Sec.~\ref{sec:crit}, the onset time $t_A^{m\,*}$ for the leading-order 
formula for the modulus, 

\begin{subequations}\label{eq:CC}
\begin{equation}\label{eq:CC:crit_modulus}
m_A(t) = f_A^{m\,c} + h_A \left(t_0 /t \right)^a\,,
\end{equation}
is smaller than $t_A^*$ by more than a factor $1000$. As a result, see the 
upper panel of Fig.~\ref{fig:BZPphiAt}, the equivalent 
formula~(\ref{eq:ML}) for the correlator,

\begin{equation}\label{eq:CC:ML}
\phi_A(t) = f_A^c + (1-f_A^c) 
\text{M}_a\left[-\left(t\omega_A^c\right)^a\right]\,,
\end{equation}\end{subequations}
describes the probing variable $A$ for all times exceeding $t = 1$ps. The 
leading-order asymptotic result for the modulus explains the response for 
$T = 251$K quantitatively within the interval $1$ps$\leqslant t \leqslant 
30$ps.

Figure~\ref{fig:BZPchi} shows that the leading-asymptotic description of 
the critical loss spectrum, Eq.~(\ref{eq:chi_power:leading}), would be 
relevant for the explanation of the loss minimum only in cases with 
$\omega_\text{min} < 2\cdot10^{-5}$ps$^{-1}$. The dashed line labeled 
$\omega^{2a}$ exhibits the asymptotic expansion for the critical loss up 
to the leading correction, Eq.~(\ref{eq:chi_power:chi}). This formula is 
relevant for $\omega < 10^{-3}$ps$^{-1}$. Even this expression is 
unsatisfactory for the discussion of the BZP results because of the large 
correction amplitude. However, the Cole-Cole spectrum describes the 
critical loss reasonably for $\omega < 0.2$ps$^{-1}$. The correction 
amplitude for the Cole-Cole law, Eq.~(\ref{eq:cccpar:cqcc}) is very small,
$\hat{K}_A^{cc} = 0.076$. The Cole-Cole law with leading correction, 
Eq.~(\ref{eq:ccc}), is shown in the upper panel of Fig.~\ref{fig:BZPchi} 
by the curve marked $ccc$, and it slightly above the leading-order result 
in the regime of large frequencies.

For the system under study, the Cole-Cole frequency reads $\omega_A^c = 
0.067$ps$^{-1}$. This frequency $\omega_A^c/(2\pi) \approx 10$GHz is small 
compared to the $1$THz scale for the normal-liquid dynamics. The value 
$-\log\omega_A^c$ is in the center of the $\log t$ interval studied by the 
results of Fig.~\ref{fig:BZP}. Therefore, the critical spectrum relevant 
for the understanding of the data differs qualitatively from the 
leading-order power-law formula. As a result, the loss minima in 
Fig.~\ref{fig:BZPchi}, which are caused by the superposition of the 
critical spectra for $\omega$ near and above $\omega_A^c$ and the 
von~Schweidler-law wing of the $\alpha$-peak, differ drastically from the 
general low-frequency shape described by Eq.~(\ref{eq:interpol}). For $T = 
251$K, the superposition yields the pseudo-von~Schweidler law manifested 
as a wing specified by the exponent $b' = 0.20$. This wing will be 
analyzed further in Sec.~\ref{sec:alphabeta}.

\subsection{Critical Relaxation for an Intermediate Cole-Cole 
Frequency\label{subsec:salol}}

\begin{figure}
\includegraphics[width=\columnwidth]{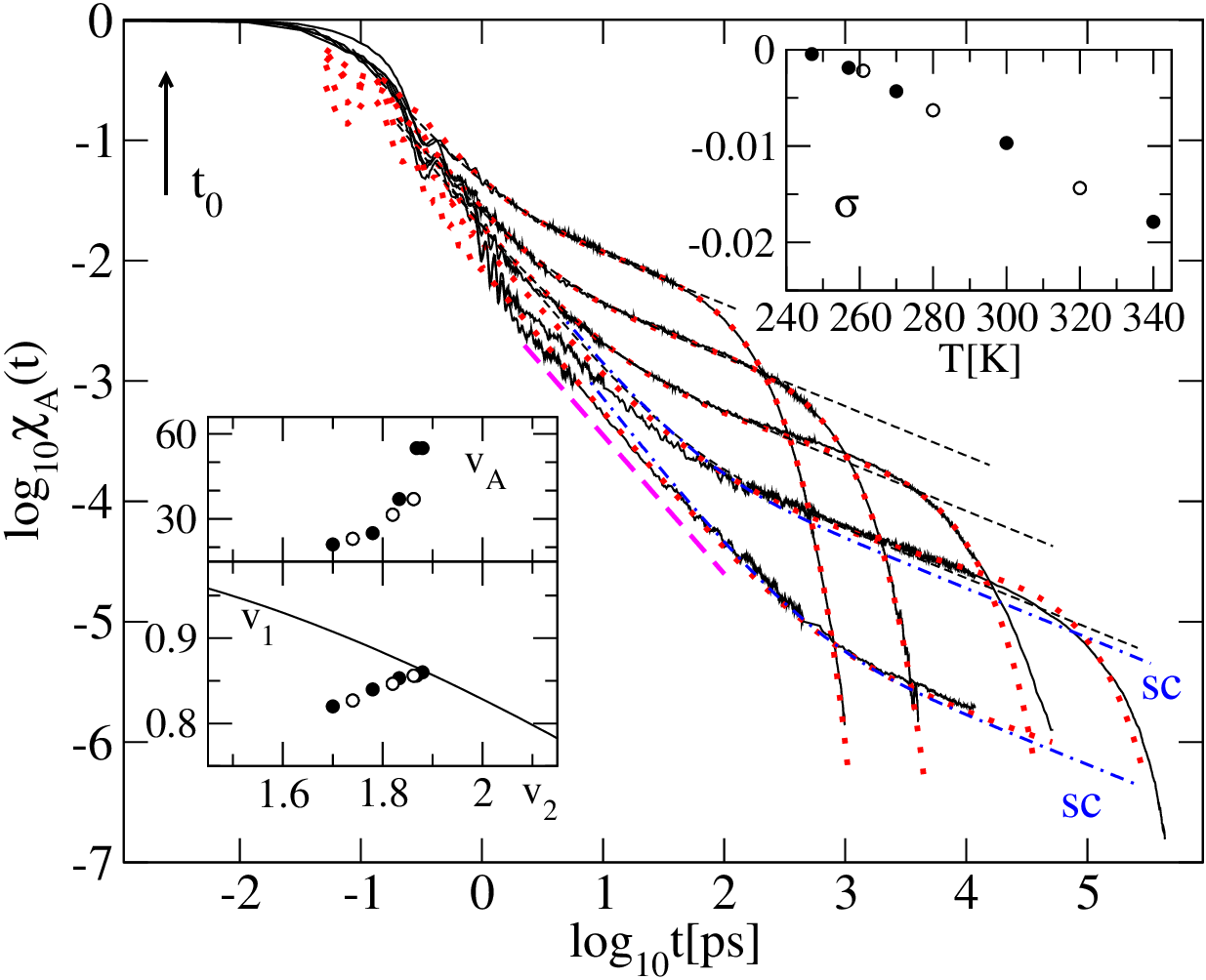}
\caption{\label{fig:salol}
OKE-response functions measured for salol for $T/K = 247,\, 257,\, 270,\, 
300,\, 340$ (full lines from bottom to top) \cite{Hinze2000b}. The dotted 
lines are fits by the schematic-model response $\chi_A(t)$ calculated with 
$\Omega = 2\Omega_A = 10\nu_A = 15.9$ps$^{-1}$ \cite{Goetze2004}. Symbols 
and curve styles are the same as in Fig.~\ref{fig:BZP}. The slope of the 
straight dashed line is $-1.15$.
}
\end{figure}

\begin{figure}
\includegraphics[width=\columnwidth]{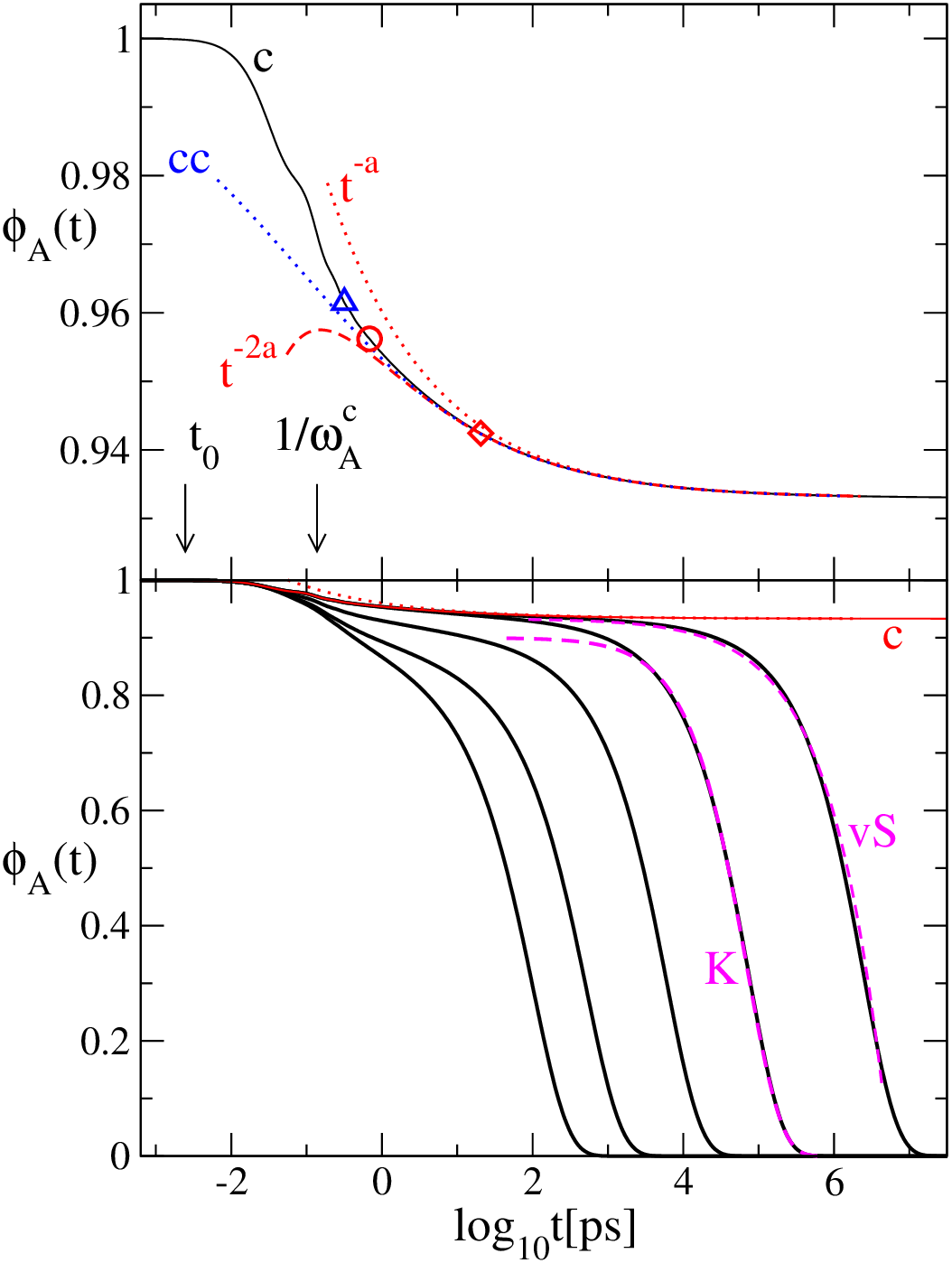}
\caption{\label{fig:salolphiAt} Correlation functions for salol.
The lower panel shows the correlators 
$\phi_A(t)$ used for the response fits in Fig.~\ref{fig:salol} for $T/K = 
247,\,257,\,270,\,300,\,$ and $340$ (heavy full lines from right to left). 
The light full line shows the critical correlator calculated for $\lambda 
= 0.73$, and $v_A = 55$. The dashed line marked \textit{vS} exhibits a 
von~Schweidler law, Eq.~(\ref{eq:vS:phi}), for a time scale chosen to 
match the $T=247$K correlator. The dashed line marked \textit{K} is a 
Kohlrausch-law fit for the $T=257$K curve with exponent $\beta = 0.95$. 
The dotted line exhibits the formula~(\ref{eq:crit:corr}) with $t_0 = 
0.00246$ps. The upper panel shows the critical decay (\textit{c}) together 
with the leading- ($t^{-a}$) and next-to-leading-order asymptotic 
solution~(\ref{eq:critcorr:phi}) ($t^{-2a}$). The dotted curve labeled 
\textit{cc} displays the leading-order solution from Eq.~(\ref{eq:ML}) 
with $\chi_0^{cc}=0.067$ and $\omega_A^c = 7.25$ps. The points of 10\% 
deviation of the approximations $t^{-a}$, $t^{-2a}$, and \textit{cc} from 
the critical decay are indicated by the diamond, circle, and triangle, 
respectively. The times $t_0$ and $1/\omega_A^c$ are marked by arrows.
}
\end{figure}

\begin{figure}
\includegraphics[width=\columnwidth]{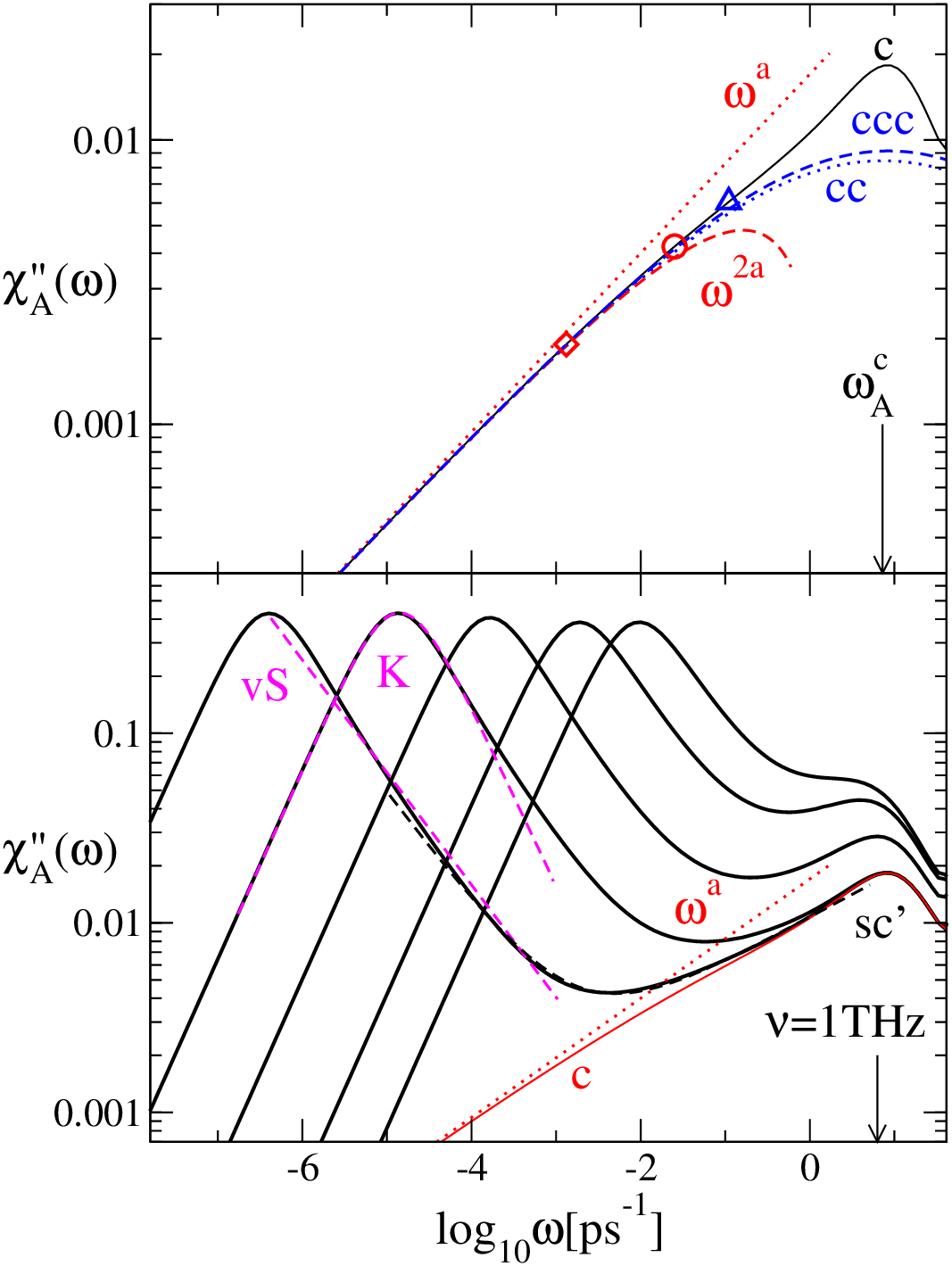}
\caption{\label{fig:salolchiAw} Spectra for salol.
The lower panel shows the susceptibility spectra $\chi_A''(\omega)$ for 
the correlators shown in the lower panel of Fig.~\ref{fig:salolphiAt}. The 
leading-order approximation to the critical spectrum, 
Eq.~(\ref{eq:chi_power:chi}), is shown as dotted straight line marked 
$\omega^a$. The dashed line labeled \textit{sc'} exhibits the 
interpolation formula (\ref{eq:interpol}) with the von~Schweidler exponent 
$b=0.59$ and the critical exponent $a$ replaced by $a'' = 0.24$. 
The upper panel exhibits the critical spectrum as full line with label 
\textit{c}, the leading-order asymptotic approximation, 
Eq.~(\ref{eq:chi_power:leading}), (dotted, labeled $\omega^a$) and 
Eq.~(\ref{eq:chi_power:chi}), (dashed, labeled $\omega^{2a}$). The 
approximation by the Cole-Cole function for $\omega_A^c = 7.25$ps$^{-1}$ 
(\textit{cc}), Eq.~(\ref{eq:ccleft:chi}), is shown dotted, the inclusion 
of the correction, Eq.~(\ref{eq:ccc}), yields the dashed curve labeled 
\textit{ccc}. The points of 10\% deviation of the approximations 
$\omega^{-a}$, $\omega^{-2a}$, and \textit{cc} from the critical spectrum 
are indicated by the diamond, circle, and triangle, respectively. The 
frequencies $\omega_A^c$ and $\omega/(2\pi) = \nu = 1$THz are marked by 
arrows.
}
\end{figure}

Five OKE response functions measured for salol \cite{Hinze2000b} and fits 
by the functions $\chi_A(t)$ of the above explained schematic model are 
reproduced in Fig.~\ref{fig:salol}. The filled dots in the insets specify 
the mode-coupling coefficients $v_1$, $v_2$, and $v_A$ used for the 
calculations. Further details can be found in Ref.~\cite{Goetze2004}, 
where, however, the value for $\Omega$ was misprinted. The series of 
states $(v_1, v_2)$ extrapolates to a transition point with $\lambda = 
0.73$, which implies a critical exponent $a=0.31$ and a von~Schweidler 
exponent $b=0.59$. The extrapolation of the $\sigma$-versus-$T$ parameters 
to $\sigma = 0$ suggests a critical temperature $T_c = 245$K with an 
estimated uncertainty of $\pm 3$K. The MCT correlators $\phi_A(t)$ and 
loss spectra $\chi''_A(\omega)$, which are equivalent to the response 
$\chi_A(t)$ in Fig.~\ref{fig:salol}, are shown in the lower panels of 
Figs.~\ref{fig:salolphiAt} and~\ref{fig:salolchiAw}, respectively. They 
can be considered as the measured results for the glassy dynamics of 
salol, since the calculated and measured curves in Fig.~\ref{fig:salol} 
agree outside the transient regime. This holds with the reservation, that 
fits and measurements for the $T = 270$K results exhibit some 
discrepancies for times exceeding 20ns. The lines with label $c$ in 
Figs.~\ref{fig:salolphiAt} and~\ref{fig:salolchiAw} show critical 
correlators and critical loss spectra, respectively, calculated for 
$\lambda = 0.73$ and $v_A = 55$. The correlators relax to the plateau 
$f_A^c = 0.93$ with a critical amplitude $h_A = 0.18$.

The test of the superposition laws for the long-time relaxation parts of 
the correlators is left to the reader. This second scaling law of MCT 
relates the $\alpha$-relaxation peaks for the loss spectra for 
$\chi''_A(\omega) \geqslant 0.1$. Kohlrausch-law fits and von~Schweidler 
asymptotes have been added to the data for $T = 257$K and $T = 247$K, 
respectively, in order to emphasize that the results for the evolution of 
the below-plateau-decay process follows the familiar pattern. Notice that 
the response for $T = 257$K exhibits the von~Schweidler-law decay, 
$\log\chi_A(t) = \text{const.} -(1-b) \log t$ for the large time interval 
0.1ns $\leqslant t \leqslant $ 10ns. The measurement for this temperature 
permits a rather precise determination of the exponent $b$ and, thereby, 
of $\lambda$.

The light-dashed lines in Fig.~\ref{fig:salol} are fits of the data by the 
first-scaling-law expression~(\ref{eq:chiA_beta}) with freely adjusted 
scales $s_\sigma$ and $t_\sigma$. For $T \geqslant 270$K, these fits 
describe the data outside the transient regime up to times within the 
von~Schweidler-law region. A more detailed documentation of this result 
can be found in Ref.~\cite{Hinze2000b}, where also consistency of the 
fitted scales $s_\sigma$, $t_\sigma$, and $t'_\sigma$ with the 
MCT-power-law formulas was demonstrated. The scaling-law results 
calculated with the MCT values for the scales $h_A$, $s_\sigma$, and 
$t_\sigma$ in Eq.~(\ref{eq:chiA_beta}) are shown in Fig.~\ref{fig:salol} 
as dash-dotted lines marked $sc$ for the $T = 257$K and $T = 247$K data. 
For $T = 257$K, $\varepsilon = -0.05$ is so large, that 
Eq.~(\ref{eq:chiA_beta}) cannot account quantitatively for the scales 
ruling the von~Schweidler-law decay. For $T \geqslant 270$K the distance 
parameter $\varepsilon$ exceeds 10\%; the leading asymptotic expansion 
formula for the plateau-crossing process is not applicable for such large 
distances of the control parameter from the critical value. In contrast, 
the scaling law~(\ref{eq:chiA_beta}) describes the $T = 247$K result well 
for the large time interval 0.04ns $<t<$ 10ns. In this case, the distance 
parameter $\varepsilon = -0.008$ is so small that the leading-order 
expansion result for the plateau crossing, Eq.~(\ref{eq:phi_gpm}), 
accounts for a major part of the long-time response -- readjusting the 
scales $(h_As_\sigma)$, and $t_\sigma$ would not improve the fit. 

Figure~\ref{fig:salolphiAt} shows that the $T = 247$K response is close to 
the critical one for $t \leqslant 1$ns. The straight dashed line with 
slope $-1.15$, which is shown in Fig.~\ref{fig:salol}, demonstrates that 
the critical response exhibits a power-law decay $\chi_A(t) \propto 
1/t^{1+a'}$, $a' = 0.15$, for 2ps $<t<$ 100ps. This is equivalent to a 
power-law decay of the correlator: $\phi_A(t) - f_a^c \propto 1/t^{a'}$. 
In qualitative agreement with the prediction by Eq.~(\ref{eq:phi_gpm}), 
the crossover from the $t^{-a}$ decay to the $-t^b$ decay causes an 
inflection point for the $\phi_A(t)$-versus-$\log t$ curve in 
Fig.~\ref{fig:salolphiAt}. The corresponding crossover from the 
$\omega^{-b}$ wing of the $\alpha$-peak to some $\omega^{a''}$ spectrum 
causes the loss minimum located near $\log\omega = -2.5$ in 
Fig.~\ref{fig:salolchiAw}. The dashed curve marked $sc'$ exhibits a 
description of this minimum by the function 
$\chi''_A(\omega)/\chi_\text{min} = \left[ b \left(\omega / 
\omega_\text{min}\right)^{a''} + a'' \left(\omega_\text{min} / 
\omega\right)^b \right]/(a''+b)$ which is suggested by 
Eq.~(\ref{eq:interpol}). For salol, as opposed to BZP, the minimum 
position decreases strongly with decreasing $|\varepsilon|$, in 
qualitative agreement with the first-scaling-law results.

The inconsistent exponents for $T = 247$K in the regime 2ps $<t<$100ps 
indicate that the leading-order asymptotic description is still not 
applicable; the exponents $a' = 0.15$ obtained from the fit of the OKE 
data and $a'' = 0.24$ obtained from the fit of the corresponding minimum
are different and both are smaller than the correct value of $a = 0.31$. 
The deviation of the critical correlator from 
the leading-order power-law result~(\ref{eq:crit:corr}) 
and~(\ref{eq:chit0}) within the specified time interval is caused by the 
large value for the correction amplitude: $\hat{K}_A = -1.78$. This value 
implies an onset time for the $t^{-a}$ decay of beyond 20ps as shown by 
the diamond in Fig.~\ref{fig:salolphiAt}. The high-frequency part of the 
loss minimum is located in the region 10ns$^{-1}<\omega<$ 1ps$^{-1}$. 
Figure~\ref{fig:salolchiAw} demonstrates that the $\omega^a$ law does not 
describe the critical loss there. Inclusion of the leading correction 
terms for the analytic description of the critical dynamics, i.e., using 
Eq.~(\ref{eq:critcorr:phi}) for the correlator and 
Eq.~(\ref{eq:chi_power:chi}) for the loss spectrum, explains the result 
for $t \geqslant 0.6$ps and $\omega \leqslant 25$ns$^{-1}$, as 
demonstrated in Figs.~\ref{fig:salolphiAt} and~\ref{fig:salolchiAw}, 
respectively.

From Eq.~(\ref{eq:cmq:cmq}), one derives the correction amplitude for the 
modulus $\hat{K}_A^m = 0.177$. The leading-order formula 
(\ref{eq:CC:crit_modulus}) for the relaxation kernel describes the kernel 
$m_A$ outside the transient regime. The corresponding expression 
(\ref{eq:CC:ML}) for the correlator and the equivalent Cole-Cole formula 
for the loss spectrum account for the critical glassy dynamics, as shown 
in Figs.~\ref{fig:salolphiAt} and~\ref{fig:salolchiAw}. The Cole-Cole 
formula with correction term of amplitude $\hat{K}_A^c = 0.09$ yields a 
slight improvement compared to the equation based on $\hat{K}_A^c = 0$, as 
shown in Fig.~\ref{fig:salolchiAw} by the curve with label $ccc$. 
Equation~(\ref{eq:cccpar:wcc}) leads to the Cole-Cole frequency for salol: 
$\omega_A^c = 7.25$ps$^{-1}$. This value is close to the loss peak for the 
normal-liquid dynamics: $\omega_A^c t_\text{mic} \approx 1$. The part of 
the glassy critical loss spectrum which is relevant for the explanation of 
the data in Fig.~\ref{fig:salol} deals with the regime $10^{-4} \leqslant 
\omega/\omega_A^c \leqslant 10^{-1}$. Within this frequency interval, the 
Cole-Cole spectrum increases smoothly with $\omega$. Therefore, the 
leading-order formulas (\ref{eq:chit0}) or (\ref{eq:chi_power:leading}) 
describe the dynamics qualitatively. However, the true critical spectrum 
differs from its low-frequency asymptote, 
Eq.~(\ref{eq:chi_power:leading}). Hence, the description of the dynamics 
by the scaling law is not correct quantitatively. Within the specified 
interval, the $\log\chi''_A(\omega)$-versus-$\log\omega$ spectrum can be 
approximated reasonably by a straight line, $\log\chi''_A(\omega) \approx 
\text{const.} + a'' \log\omega$, $a'' = 0.24$. As a result, the critical 
dynamics is approximated well by a power law decay specified by an 
exponent $a''$ smaller than the critical exponent $a$.

In the preceding section, the critical spectra have been discussed for 
both small and intermediate Cole-Cole frequencies $\omega_q^c$; the case 
of a large Cole-Cole frequency is found in the 
depolarized-light-scattering spectra for CKN \cite{Li1992,Sperl2005bpre}. 
In the fits with a schematic model by Alba-Simionesco and coworkers 
\cite{Krakoviack1997,Krakoviack2002} one can identify a Cole-Cole peak for 
the critical spectra, but the description with the power-law solution is 
superior if terms up to order $\omega^{2a}$ are considered.

\section{Cole-Cole wing\label{sec:alphabeta}}

Figures~\ref{fig:BZP}--\ref{fig:BZPchi} and 
\ref{fig:salol}--\ref{fig:salolchiAw} demonstrate the scenarios for the 
evolution of the glassy dynamics for $\omega^c_q \ll t^{-1}_\text{mic}$ 
and $\omega^c_q \approx t^{-1}_\text{mic}$, respectively, for the probing 
variable described by the second correlator of a two-component schematic 
MCT model. Solutions for this model, which exemplify a scenario as shown 
in Fig.~\ref{fig:BZPchi}, have been discussed before by Cummins 
\cite{Cummins2005}. He pointed out that the evolution of the wing 
phenomenon obtained from the model is similar to the one known for 
dielectric-loss spectroscopy for glycerol and several van-der-Waals 
liquids \cite{Dixon1990}. He emphasizes also that the cited measurements 
refer to temperatures $T$ below $T_c$, while the calculations are done for 
$T > T_c$. Furthermore, the experimental data demonstrate the wing only 
for frequencies below 10GHz$\approx 10^{-2}t^{-1}_\text{mic}$. So far, no 
wing has been reported which occurs in the two-decade frequency window 
adjacent to the microscopic excitation region.

In order to describe the wing phenomenon quantitatively, one can extend 
the Cole-Cole formula so that the spectrum can be described also for small 
but non-vanishing separation parameters $\sigma = C \varepsilon \propto 
(T_c-T)/T_c$. The leading-order expression for the plateau-crossing 
process of the fluctuating-force correlators reads in analogy to 
Eq.~(\ref{eq:phi_gpm}): $m_q(t) = f_q^{m\,c} + h_q^m 
g_\pm(t/t_\sigma),\,\varepsilon\gtrless 0$. Laplace transformation yields 
$\omega m_q(\omega) = -f_q^{m\,c} + h_q^m C(\omega)$. Function $C(\omega)$ 
obeys a scaling law,

\begin{equation}\label{eq:bet_C}
C(\omega) = \sqrt{|\sigma|} c_\pm(\omega t_\sigma)\,,\quad 
\varepsilon \gtrless 0\,,
\end{equation}
with the control-parameter independent shape functions $c_\pm(\hat{\omega})$ 
given by the Laplace transforms $g_\pm(\hat{\omega})$ of the shape 
functions $g_\pm(\hat{t})$:
$c_\pm (\hat{\omega}) = \hat{\omega} g_\pm (\hat{\omega})$.
Substitution of $\omega m_q(\omega)$ into Eq.~(\ref{eq:MCTlong:chi}) 
yields the desired result. Expansion of the right-hand side in terms of 
the small parameter $C(\omega)$ and comparison with the result for 
$\chi_q(\omega)$ following from Eq.~(\ref{eq:phi_gpm}) reproduces the 
relations (\ref{eq:fqmc},\ref{eq:hmq}) for the critical arrested parts and 
critical amplitudes. One gets

\begin{subequations}\label{eq:ccbeta}
\begin{equation}\label{eq:ccbeta_lead}
\chi_q(\omega) = \left( 1- f_q^c \right) /
\left\{ 1 - \left[h_q/(1-f_q^c)\right] C(\omega)\right\}\,.
\end{equation}
Equation~(\ref{eq:ccbeta_lead}) was found for the schematic model in 
Ref.~\cite{Goetze1989c} for the special case of the second correlator 
$\phi_A(t)$ and in the limit of infinite coupling strength $v_A$. It is 
shown here, that Eq.~(\ref{eq:ccbeta_lead}) can be derived without such 
restrictions. The leading correction to Eq.~(\ref{eq:ccbeta_lead}) can be 
calculated from the asymptotic expansions in \cite{Franosch1997,Fuchs1998} 
in a straightforward manner; however, the evaluation of the result for 
fitting data is rather involved. Focusing on the critical decay, one can 
extend the preceding formula by adding the large-frequency part of the 
correction:

\begin{equation}\label{eq:ccbeta_corr}\begin{split}
\chi_q(\omega) = \left( 1- f_q^c \right) /
\left\{ 1 - \left[h_q/(1-f_q^c)\right] C(\omega)\right.\\\left.
+ \hat{K}_q^\text{cc}\left(-i\omega/\omega_q^c\right)^{2a}\right\}\,.
\end{split}\end{equation}
\end{subequations}
This expression describes the small-$\varepsilon$ dynamics for $\omega 
t_\text{mic}<1$ and frequencies extending down to the beginning of the 
von~Schweidler-law decay. 

\begin{figure}[ht]
\includegraphics[width=\columnwidth]{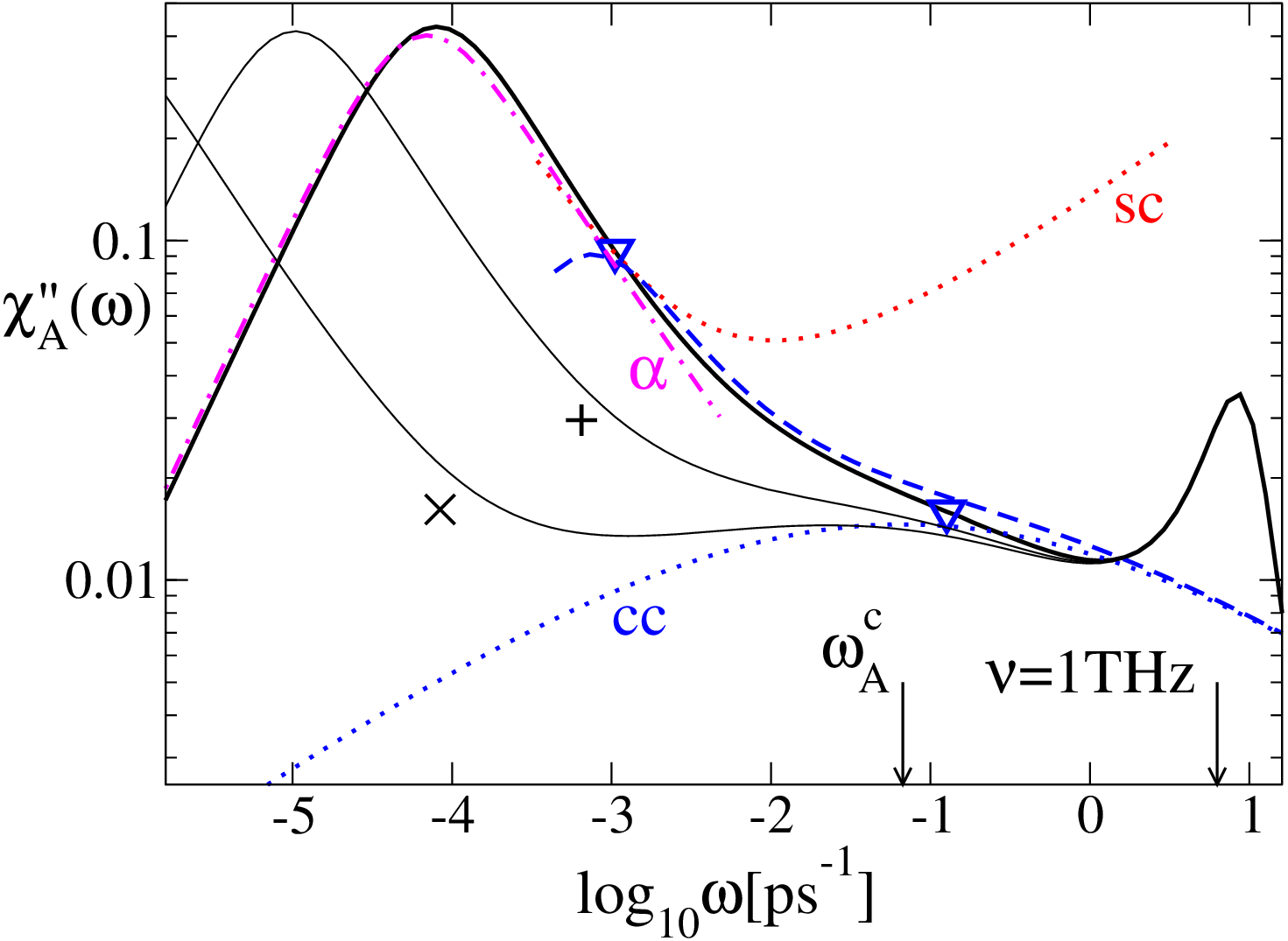}
\caption{\label{fig:BZPwing}Asymptotic description of the wing in BZP.
The heavy full line reproduces the BZP loss spectrum for $T=251$K from 
Fig.~\ref{fig:BZPchi} and the dashed one exhibits the spectrum of the 
extended Cole-Cole susceptibility, Eq.~(\ref{eq:ccbeta_lead}). The 
triangles mark the points of 10\% deviation between these spectra. 
The dotted line \textit{sc} is the first-scaling-law result for the 
susceptibility, and the line \textit{cc} shows the Cole-Cole spectrum for 
the critical point. The dash-dotted line represents the 
second-scaling-law result for the loss peak. The thin full lines are 
solutions corresponding to the state points indicated by $+$ and $\times$ 
in Fig.~\ref{fig:BZP}.
}
\end{figure}

BZP exemplifies the case of such a small $\hat{K}^{cc}_q$ that the 
leading-order result for the modulus, Eq.~(\ref{eq:ccbeta_lead}), can be 
used. Figure~\ref{fig:BZPwing} demonstrates that this result accounts for 
the $T=251$K spectrum for the large dynamical range $10^{-3}\text{ps}^{-1} 
< \omega < 1\text{ps}^{-1}$. The scaling-law result for the loss spectrum, 
$\chi_A''(\omega) = h_A C_A''(\omega)$, is shown as dotted line 
\textit{sc}; it can describe only a part of the von~Schweidler-law 
spectrum for $\omega < 10^{-3}\text{ps}^{-1}$, but is inadequate for 
larger frequencies. This observation for the dynamics in the frequency 
domain is equivalent to the one demonstrated in Fig.~\ref{fig:BZP} for the 
results in the time domain. The second-scaling-law result for the loss 
spectrum is exhibited as dash-dotted line. It follows from 
Eq.~(\ref{eq:alpha}), $\chi_A''(\omega) = (\omega t_\sigma') 
\tilde{\phi}_A''(\omega t_\sigma')$, and accounts for the loss peak for 
$\omega < 10^{-3}\text{ps}^{-1}$. Combining the result for the second 
scaling law for the modulus with the result for the first scaling law for 
the loss spectrum explains the BZP spectrum for $T = 251$K for $\omega < 
1\text{ps}^{-1}$ including in particular the $\alpha$-peak wing. The small 
but systematic discrepancies between the loss spectrum in the 
structural-relaxation regime and the asymptotic MCT expressions result 
from the fact that the distance parameter $\varepsilon = (T-T_c)/T_c 
\approx 0.07$ is so large, that the asymptotic results for the scales 
still have noticeable errors while the shape functions already agree well 
with the nontrivial spectra.

Starting from the state that fits the 251K data for BZP, a number of 
extrapolations are possible within the schematic model. Keeping all 
parameters but $(v_1, v_2)$ fixed, two additional states shall be 
considered; they are indicated by $\times$ and $+$ in the left inset of 
Fig.~\ref{fig:BZP}; and their spectra are marked accordingly in 
Fig.~\ref{fig:BZPwing}. The first state point ($\times$) is close to the 
transition point and is characterized by $\sigma = -0.003$ ($\varepsilon 
=-0.01$). Solutions at this point are shown in Ref.~\cite{Goetze2004} for 
$\chi(t)$ and $\chi_A(t)$. The respective spectrum $\chi_A''(\omega)$ in 
Fig.~\ref{fig:BZPwing} exhibits nearly-constant loss for $0.4\, 
10^{-3}$ps$^{-1} < \omega < 0.4$ps$^{-1}$ within a 10\% margin. In this 
frequency window the maximum around $\omega^c_A$ indicates an emerging 
Cole-Cole peak; the minimum at $\omega = 10^{-3}$ps$^{-1}$ is caused by 
the crossover of the high-frequency wing of the $\alpha$-peak and the 
low-frequency wing of the Cole-Cole peak. This minimum is ruled by the 
first scaling law and the divergent time scale $t_\sigma$. The second 
state ($+$) is an interpolation between $\times$ and the solution for BZP; 
the separation parameter is $\sigma = -0.01$ ($\varepsilon = -0.03$), and 
the point is chosen similar to the ones used in Fig.~4 of 
Ref.~\cite{Cummins2005}. The spectrum in Fig.~\ref{fig:BZPwing} shows a 
wing, $\chi_A''(\omega) \propto \omega^{-0.1}$, for almost three orders of 
magnitude in frequency. Hence, the crossover between $\alpha$- and 
Cole-Cole peaks can be interpreted as a power-law wing $\omega^{-b'}$ for 
some range in frequencies and control parameters without fine-tuning. The 
evolution of the crossover progresses from a wing with gradually lower 
exponents $b'$ to nearly-constant loss and eventually the emergence of a 
Cole-Cole peak. While the extrapolations presented in this paragraph are 
not at all implied by the fit of the data, the scenarios seem nevertheless 
possible for realistic parameter values.

\begin{figure}[ht]
\includegraphics[width=\columnwidth]{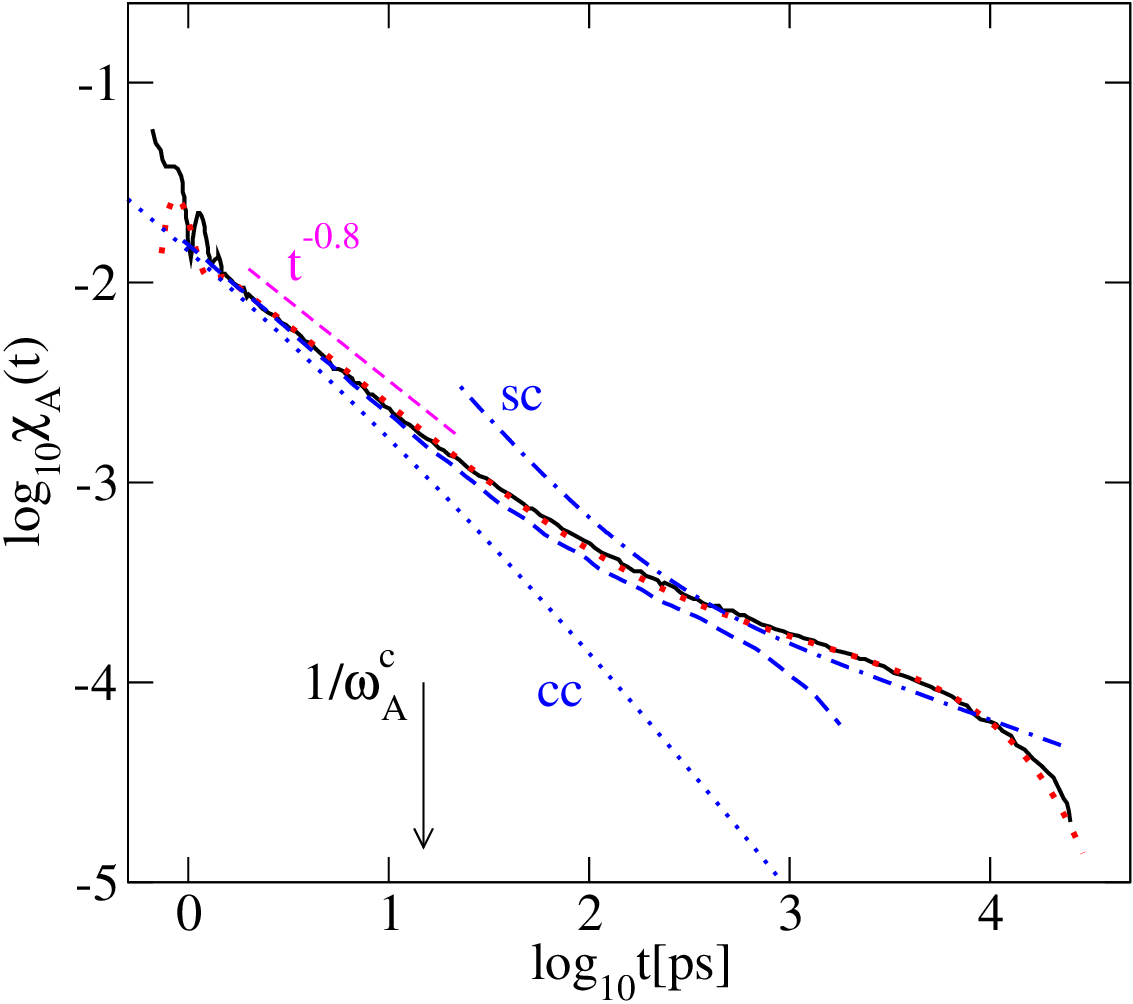}
\caption{\label{fig:BZPOKE}Asymptotic description of the OKE data for BZP.
The full line reproduces the BZP data for $T=251$K from \cite{Cang2003c}; 
the dotted line reproduces the fit from Fig.~\ref{fig:BZP}. The dashed 
curve, the dash-dotted curve \textit{sc}, and the dotted curve \textit{cc} 
respectively show the Fourier backtransforms of the extended Cole-Cole 
susceptibility in Eq.~(\ref{eq:ccbeta_lead}), of the first scaling law, 
and of the Cole-Cole law in Eq.~(\ref{eq:ccleft:chi}), cf. 
Fig.~\ref{fig:BZPwing}.
}
\end{figure}

To conclude the discussion of BZP, let us return to the OKE data for 
$T=251$K as shown in Fig.~\ref{fig:BZPOKE}. In addition to data and fit 
for this state, the Fourier backtransforms of the asymptotic curves from 
Fig.~\ref{fig:BZPwing} are displayed. While the asymptotic solution at the 
critical point (labeled \textit{cc}) covers only less than a decade, the 
extended Cole-Cole susceptibility in Eq.~(\ref{eq:ccbeta_lead}) describes 
the data over three orders of magnitude in time and thereby explains the 
$t^{-0.8}$-law found empirically. The remaining discrepancies are again 
due to the relatively large distance parameter for this state; Fig.~3 of 
Ref.~\cite{Goetze2004} shows prefect agreement with the Cole-Cole law for 
states closer to the transition including state $\times$ from 
Fig.~\ref{fig:BZP}.

\begin{figure}[ht]
\includegraphics[width=\columnwidth]{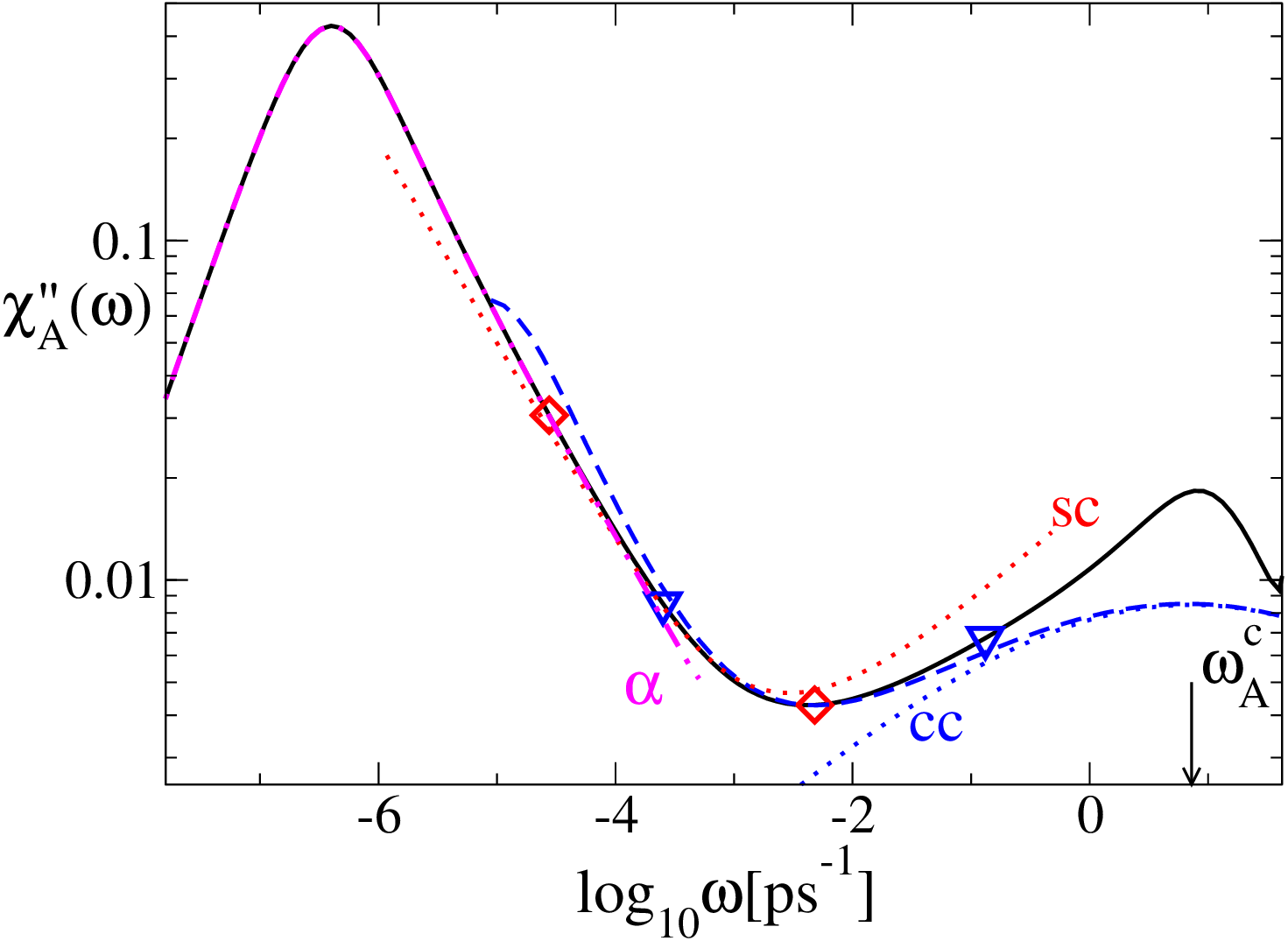}
\caption{\label{fig:Salolwing}Asymptotic description of the minimum in 
salol. The full line reproduces the salol loss spectrum for $T=247$K from 
Fig.~\ref{fig:salolchiAw} and the dashed one exhibits the spectrum of the 
extended Cole-Cole susceptibility, Eq.~(\ref{eq:ccbeta_lead}). The 
triangles mark the points of 10\% deviation between these spectra. 
The dotted line  \textit{cc} is the Cole-Cole spectrum for the critical 
point, and the line \textit{sc} shows the first-scaling-law result for the 
susceptibility; the diamonds mark the points where \textit{sc} deviates by  
10\% from the loss spectrum. The dash-dotted line represents the 
second-scaling-law result for the loss peak.
}
\end{figure}

Different from the case for BZP, the lowest temperature available for 
salol, $T = 247$K, is described by a much smaller distance parameter, 
$\varepsilon \approx 0.01$, and therefore shares a larger frequency 
regime, say $\omega\lesssim 0.1$ps$^{-1}$, with the critical loss 
spectrum, see the lower panel of Fig.~\ref{fig:salolchiAw}. Being so close 
to the critical point, the second scaling law in Fig.~\ref{fig:Salolwing} 
-- labeled $\alpha$ -- is indistinguishable from the loss spectrum for 
$\omega \leqslant 10^{-4}$ps$^{-1}$. The first scaling law \textit{sc} 
describes the solution up to around $\omega\lesssim 0.005$ps$^{-1}$ which 
is the same as for the solution at the critical point, cf. upper panel of 
Fig.~\ref{fig:salolchiAw}. As for the critical spectrum, the leading-order 
approximation in Eq.~(\ref{eq:ccbeta_lead}) improves the description to 
larger frequencies by two decades up to $\omega\approx 0.1$ps$^{-1}$ 
before crossing over to the Cole-Cole peak which in this case is hidden 
under the microscopic excitations.

\section{Conclusion\label{sec:sum}}

Within the regime of glassy dynamics, susceptibilities $\chi_q(\omega)$ 
and moduli $\omega m_q(\omega)$ are related by Eq.~(\ref{eq:MCTlong:chi}); 
and at the critical point, susceptibilities and moduli are both described 
by the universal power law $\omega^a$. The range of validity of this power 
law is given by amplitudes for the correction $\omega^{2a}$ -- $\hat{K}_q$ 
for the susceptibilities and $\hat{K}_q^m$ for the moduli, which are 
related by Eq.~(\ref{eq:cmq:cmq}). For some parameter regions a large 
value of $|\hat{K}_q|$ can render the power-law expansion for the 
susceptibilities irrelevant, while the correction amplitude 
$|\hat{K}_q^m|$ is so small that the universal power-law can be used for 
the description of the modulus successfully. Typically, this occurs if 
$\hat{K}_q$ is large and negative, and this can be expected if the plateau 
value $f^c_q$ is high, cf. Eq.~(\ref{eq:cmq:cmq}). In this case, the 
susceptibility can be described well by the expansion of its inverse, 
Eq.~(\ref{eq:ccc}). For vanishing correction $\hat{K}_q^m = K_q^{cc} = 0$, 
formula, the susceptibility at the critical point is given by the 
Cole-Cole law~(\ref{eq:ccleft:chi}), a formula first introduced in 1941 
\cite{Cole1941}.

The Cole-Cole frequency $\omega_q^c$ in Eq.~(\ref{eq:ccleft:chi}) 
introduces a characteristic scale for assorting the critical dynamics into 
three categories. (1) If $\omega_q^c$ is large compared to the scale 
$t^{-1}_\text{mic}$ for the band of normal liquid excitations, the 
Cole-Cole susceptibility reduces to the leading-order scaling-law formula 
which is implied by Eq.~(\ref{eq:phi_gpm}). There is a control-parameter 
sensitive minimum as discussed for Eq.~(\ref{eq:interpol}) that 
interpolates between the von~Schweidler-law tail of the $\alpha$-peak, 
$\chi_q''(\omega) \propto 1/\omega^b$, and the critical spectrum 
$\chi_q''(\omega) \propto \omega^a$, where $a$ and $b$ are related via the 
exponent parameter $\lambda$. As an example for such a scenario one can 
cite the molten salt CKN. While the leading-order result alone is not 
sufficient to describe the measured data quantitatively, the power-law 
expansion in Eq.~(\ref{eq:chi_power:chi}) yields a satisfactory 
description that is superior to the Cole-Cole solution 
\cite{Sperl2005bpre}. (2) If $\omega^c_q$ is close to $t^{-1}_\text{mic}$, 
one encounters a scenario demonstrated in Fig.~\ref{fig:Salolwing} for 
salol. There is a control-parameter sensitive loss minimum. It originates 
from the crossover between the von~Schweidler-law tail and the critical 
spectrum; and the critical spectrum approaches the maximum of the 
underlying Cole-Cole peak. A description by Eq.~(\ref{eq:interpol}) is 
possible only, if the exponent $a$ is replaced by some effective one 
$a''$, which is smaller than $a$, cf. Fig.~\ref{fig:salolchiAw}. (3) If 
$\omega_q^c$ is smaller than the microscopic time scale, say $\omega_q^c 
t_\text{mic} \approx 0.05$, we obtain the scenario seen in 
Fig.~\ref{fig:BZPwing} for benzophenone (BZP). In this case, the 
von~Schweidler-law part of the $\alpha$-peak, $\chi_q'' \propto 
1/\omega^b$, crosses over to some flatter wing, $\chi_q'' \propto 
1/\omega^{b'}$, $a < b' < b$. This wing is caused by approaching the 
Cole-Cole spectrum for frequencies around the maximum at $\omega^c_q$. For 
yet higher frequencies, one encounters a crossover -- from the 
high-frequency part of the Cole-Cole spectrum for structural relaxation to 
the spectrum due to normal-liquid excitations -- producing a different 
kind of a minimum. The position of this minimum is control-parameter 
insensitive. For the understanding of this minimum, both the scaling-law 
(\ref{eq:chiA_beta}) and the interpolation formula (\ref{eq:interpol}) are 
irrelevant. On the other hand, as shown in Sec.~\ref{sec:alphabeta} and 
especially in Fig.~\ref{fig:BZPwing}, it is possible to describe the wing 
rather accurately by the leading-order formula~(\ref{eq:ccbeta_lead}). If 
the $\alpha$-peak is shifted to yet lower frequencies, the wing can give 
way to a separate Cole-Cole peak.

All figures in this work have been prepared with model parameters that 
reproduce the OKE-response functions of benzophenone (BZP) and salol. 
Hence, Fig.~\ref{fig:BZPwing} implies that the BZP spectra for 
temperatures near 250K, when measured by depolarized light-scattering in 
backward direction, should exhibit an $\alpha$-peak wing for frequencies 
$\nu$ between about 1GHz and about 100GHz. The crossover from the 
von~Schweidler-law wing to the wing induced by the Cole-Cole peak is 
expected to occur around 1GHz. This crossover position depends sensitively 
on the temperature, because the von~Schweidler-law relaxation depends 
sensitively on $T$.

The Cole-Cole law with correction, Eq.~(\ref{eq:ccc}), has been derived in 
its fully general microscopic version in Sec.~\ref{sec:crit}; only in 
Secs.~\ref{sec:appl} and~\ref{sec:alphabeta} the theory was specialized to 
schematic models. While for the schematic models a number of parameters 
can be fixed to describe experimental data, for microscopic models the 
static structure of the model system determines these parameters uniquely. 
A first example for Cole-Cole dynamics in a microscopic model has been 
discussed recently for the mean-squared displacement $\delta r^2(t)$ of 
the hard-sphere system \cite{Sperl2005}; including the Mittag-Leffler 
function for the critical relaxation allows for an analytic description of 
the mean-squared displacement for the full range of the dynamics similar 
to BZP in Fig.~\ref{fig:BZPwing}. In addition, the Cole-Cole dynamics was 
identified in the data measured by van~Megen \textit{et. al} 
\cite{Megen1998} where Eq.~(\ref{eq:ccc}) accounts for the data for an 
interval in time of three orders of magnitude adjacent to the transient 
dynamics. Hence, the results discussed above seem relevant for both 
molecular and colloidal glasses and can be expected to facilitate more 
detailed investigations.

I want to thank W.~G\"otze for repeated help during the preparation of 
this work; I am also grateful to L. Berthier, H.~Z.~Cummins, E.~R\"o\ss 
ler, and Th.~Voigtmann for enlightening discussions. Support was provided 
by DFG grant No. SP~714/3-1 and NSF grants DMR0137119 and DMS0244492.

\bibliographystyle{apsrev}
\bibliography{lit,add}

\end{document}